\documentclass[11pt]{article}

\usepackage{amssymb,amsfonts,amsmath,amsthm,amscd,dsfont,mathrsfs}
\usepackage{graphicx,float,psfrag,epsfig}
\usepackage{wrapfig}

\DeclareMathAlphabet{\mathpzc}{OT1}{pzc}{m}{it}

\newtheorem{propo}{Proposition}[section]
\newtheorem{lemma}{Lemma}
\newtheorem{definition}[propo]{Definition}
\newtheorem{coro}[propo]{Corollary}
\newtheorem{thm}[propo]{Theorem}
\newtheorem{theorem}[propo]{Theorem}

\newtheorem{remark}[propo]{Remark}

\def\tdelta{{\tilde \delta}}
\def\tP{{\tilde P}}
\def\cR{{\cal R}}

\def\cM{{\cal M}}
\def\cD{{\cal D}}

\def\cX{{\cal X}}

\def\tX{{\tilde X}}
\def\tx{{\tilde x}}
\def\tM{{\tilde M}}
\def\cY{{\cal Y}}

\def\cA{{\cal A}}
\def\bS{{\bar S}}

\def\tdelta{{\tilde \delta}}
\def\tepsilon{{\tilde \varepsilon}}

\def\Z{{\mathbb Z}}
\def\tL{{\tilde L}}
\def\ones{{\mathds 1}}

\def\PFA{P_{\rm FA}}
\def\PMD{P_{\rm MD}}

\def\ind{{\mathbb I}}
\def\E{{\mathbb E}}
\def\prob{{\mathbb P}}
\def\reals{{\mathbb R}}
\def\conv{{\rm conv}}
\def\compose{\text{{\sc Compose($\cA,\cM,k,b$)}}}

\newcommand{\cT}{{\cal T}}
\newcommand{\hf}{{\hat{f}}}

\newcommand{\aveacc}{{\rm ACC_{ave}}}
\newcommand{\wcacc}{{\rm ACC_{wc}}}
\newcommand{\w}{w}

\begin{document}

\title{The Composition Theorem for Differential Privacy}

\author{
 Peter Kairouz\footnote{
Department of Electrical and Computer Engineering, University of Illinois at Urbana-Champaign,
Email: kairouz2@illinois.edu} { }
Sewoong Oh\footnote{Department of Industrial and Enterprise Systems Engineering, University of Illinois at Urbana-Champaign,
Email: swoh@illinois.edu } { }
 Pramod Viswanath\footnote{
Department of Electrical and Computer Engineering, University of Illinois at Urbana-Champaign,
Email: pramodv@illinois.edu}
\footnote{This paper was presented in part at 2015 International Conference on Machine Learning \cite{kov14-0} and  the Twenty-ninth Annual Conference on Neural Information Processing Systems in 2015 \cite{KOV15}. }
}

\date{}

\maketitle

\begin{abstract}
Sequential querying of differentially private mechanisms degrades the overall privacy level. In this paper, we answer the fundamental question of characterizing the level of overall privacy degradation as a function of the number of queries and the privacy levels maintained by each privatization mechanism. Our solution is complete: we prove an upper bound on the overall privacy level and construct a sequence of privatization mechanisms that achieves this bound. The key innovation is the introduction of an operational interpretation of differential privacy (involving hypothesis testing) and the use of new data processing inequalities. Our result improves over the state-of-the-art, and has immediate applications in  several problems studied in the literature including differentially private multi-party computation. 
\end{abstract}

\section{Introduction}
\label{sec:int}
Differential privacy is a formal framework to quantify to what extent individual privacy in a statistical database is preserved
while releasing useful aggregate information about the database. It provides strong privacy guarantees by requiring the
indistinguishability of whether an individual is in the database or not based on the released information, regardless of the side information on the other aspects of the database the adversary may possess. Denoting the database when the individual is present as $D$ and as $D'$ when the individual is not, a differentially private mechanism provides indistinguishability guarantees with respect to the pair $(D,D')$. More generally, we consider pairs of databases that indistinguishability is guaranteed for as ``neighbors".   The formal definition of $(\epsilon,\delta)$-differential privacy is the following.
\begin{definition}[Differential Privacy \cite{DMNS06,DKM06}]
	A randomized mechanism $M$ over a set of databases is {\em $(\varepsilon,\delta)$-differentially private}
	if for all pairs of neighboring databases $D$ and $D'$, and for all  sets
	$S$ in the output space of the mechanism $\cX$,
	\begin{eqnarray*}
		\prob(M(D)\in S) &\leq& e^{\varepsilon} \,\prob(M(D')\in S) + \delta\;.
	\end{eqnarray*}
\end{definition}
A basic problem in differential privacy is how privacy of a fixed pair of neighbors $(D,D')$ degrades under {\em composition} of interactive queries  when each query, individually, meets certain differential privacy guarantees. A routine argument shows that the composition of $k$ queries, each of which is $(\epsilon,\delta)$-differentially private, is at least $(k\epsilon,k\delta)$-differentially private \cite{DMNS06,DKM06,DL09,DRV10}. A tighter bound of
$(\tepsilon_\tdelta,k\delta+\tdelta)$-differential privacy under $k$-fold adaptive composition
is provided, using more sophisticated arguments, in \cite{DRV10} for the case when each of the individual queries is $(\epsilon,\delta)$-differentially private. Here $\tepsilon_\tdelta = O\Big(k\varepsilon^2+\varepsilon\sqrt{k\log(1/\tdelta)}\Big)$. On the other hand, it was not known if this bound could be improved until this work.


Our main result is the {\em exact} characterization of the privacy guarantee  under $k$-fold composition.
Any $k$-fold adaptive composition of  $(\varepsilon,\delta)$-differentially private mechanisms
satisfies this privacy guarantee, stated as Theorem~\ref{thm:composition}.
Further, we demonstrate a specific sequence of privacy mechanisms which under (in fact, nonadaptive) composition actually degrade privacy to the level guaranteed. Our result entails a strict   improvement over the state-of-the-art: this can be seen immediately in the following approximation -- using the same notation as above, the value of $\tepsilon_\tdelta$ is now reduced to $\tepsilon_\tdelta = O\Big(k\varepsilon^2 + \varepsilon \sqrt{k\log(e+(\varepsilon\sqrt{k} /\tdelta)\,)} \,\Big)$. Since a typical choice of $\tdelta$ is $\tdelta=\Theta(k\delta)$,
in the regime where $\varepsilon=\Theta(\sqrt{k} \delta)$, this improves the existing guarantee by a logarithmic factor. The gain is especially significant when both $\varepsilon$ and $\delta$ are small.

We  start with the view of differential privacy as providing certain guarantees for the two error types (false alarm and missed detection) in a binary hypothesis testing problem (involving two neighboring databases), as in previous work \cite{WZ10}.  We  brings two benefits  of this {\em operational} interpretation of the privacy definition to bear on the problem at hand.
 \begin{itemize}
 \item The first is conceptual: the operational setting directs the logic of the steps of the proof, makes the arguments straightforward and readily allows generalizations such as heterogeneous compositions.

 \item The second is technical: the operational interpretation of hypothesis testing brings both the natural data processing inequality, and the strong converse to the data processing inequality. These inequalities, while simple by themselves, lead to surprisingly strong technical results. As an aside, we mention that there is a strong tradition of such derivations in the information theory literature: the Fisher information inequality \cite{Bla65,Z98}, the entropy power inequality \cite{Sta59,Bla65,VG06}, an extremal inequality involving mutual informations \cite{LV07},  matrix determinant inequalities \cite{CT92}, the Brunn-Minkowski inequality and its functional analytic variants \cite{DCT91} -- Chapter~17 of \cite{CT06} enumerates a  detailed list -- were all derived using operational interpretations of mutual information and corresponding data processing inequalities.

 \end{itemize}

One special case of our results, the strengthening of the state-of-the-art result in \cite{DRV10},  could also have been arrived at directly by using stronger technical methods than used in \cite{DRV10}. Specifically, we use a direct expression for the privacy region (instead of an upper bound) to arrive at our strengthened result.

The optimal composition theorem (Theorem \ref{thm:composition})
provides a fundamental limit on how much privacy degrades
under composition.
Such a characterization is a basic result in differential privacy and
has been used widely in the literature \cite{DRV10,HLM10,BBDS12,GRU12,MN12,HR13}.
In each of these instances, the optimal composition
theorem derived here (or the simpler characterization of Theorem \ref{thm:closedform})
could be ``cut-and-pasted",
allowing for corresponding strengthening of their conclusions.
We demonstrate this strengthening for two instances: variance of noise adding mechanisms in Section~\ref{sec:variance}
and \cite{BBDS12}
in Appendix~\ref{sec:Application1}.
We further show that a variety of existing noise adding mechanisms ensure the same level of
privacy with similar variances.
This implies that there is nothing special about the popular choice of adding a Gaussian noise when composing multiple queries, and the same utility as measured through the noise variance can be obtained using other known mechanisms. As an application to the operational definition of differential privacy, we prove, in Section \ref{sec:multiparty}, that a simple non-interactive randomize response mechanism is optimal in secure multi-party computation.
We start our discussions by operationally
introducing  differential privacy  as certain guarantees on the error probabilities in a binary hypothesis testing problem.

%
%

\section{Differential Privacy as Hypothesis Testing}

	Given a  random output $Y$
		of a database access mechanism $M$,
		consider the following hypothesis testing experiment.
		We choose a null hypothesis as database $D_0$
		and alternative hypothesis as $D_1$:
		\begin{eqnarray*}
		H0&:& Y \text{ came from a database } D_0\;, \\
		H1&:& Y \text{ came from a database } D_1\;.
		\end{eqnarray*}
		For a choice of a rejection region $S$,
		the probability of false alarm (type I error), when the null hypothesis is true but rejected,
		is defined as
		$\PFA(D_0,D_1,M,S) \equiv \prob\big( M(D_0)\in S\big)$, and
		the probability of missed detection (type II error), when the null hypothesis is false but retained,
		is defined as
		$\PMD(D_0,D_1,M,S) \equiv \prob\big(M(D_1) \in \bS\big)$
		where $\bS$ is the complement of $S$.
		The differential privacy condition on
		a mechanism $M$ is equivalent to
		the following  set of constraints on
		the probability of false alarm and missed detection.
		Wasserman and Zhu
		proved that ($\varepsilon,0$)-differential privacy implies
		the conditions \eqref{eq:hypo} for
		a special case when $\delta=0$ \cite[Theorem 2.4]{WZ10}.
		The same proof technique can be used to prove a
		similar result for general $\delta\in[0,1]$, and
		to prove that the conditions \eqref{eq:hypo}
		imply ($\varepsilon,\delta$)-differential privacy as well.
		We refer to Section \ref{sec:hypo} for a proof.
\begin{thm}	
		\label{thm:hypo}
		For any $\varepsilon\geq0$ and $\delta\in[0,1]$,
		a database mechanism $M$ is
		$(\varepsilon,\delta)$-differentially private
		if and only if
		the following conditions are satisfied for
		all pairs of neighboring databases $D_0$ and  $D_1$,
		and all rejection region $S\subseteq\cX$:
		\begin{eqnarray}
			\PFA(D_0,D_1,M,S) + e^\varepsilon \PMD(D_0,D_1,M,S) &\geq& 1-\delta \;, \text{ and }
			 \label{eq:hypo}\\
			e^\varepsilon\PFA(D_0,D_1,M,S) + \PMD(D_0,D_1,M,S)  &\geq& 1-\delta \;.\nonumber
		\end{eqnarray}
\end{thm}
	This operational perspective of differential privacy relates the
	privacy parameters $\varepsilon$ and $\delta$ to
	a set of conditions on probability of false alarm and missed detection.
	This 	shows that it is impossible to get both small $\PMD$ and $\PFA$
	from data obtained via a differentially private mechanism, and that the converse is also true.
	This operational interpretation of differential privacy
	suggests a graphical representation of differential privacy as illustrated in Figure~\ref{fig:region1}.
	We define the {\em privacy region} for $(\varepsilon,\delta)$-differential privacy as
	\begin{figure}[t]
	\begin{center}
		\includegraphics[width=.4\textwidth]{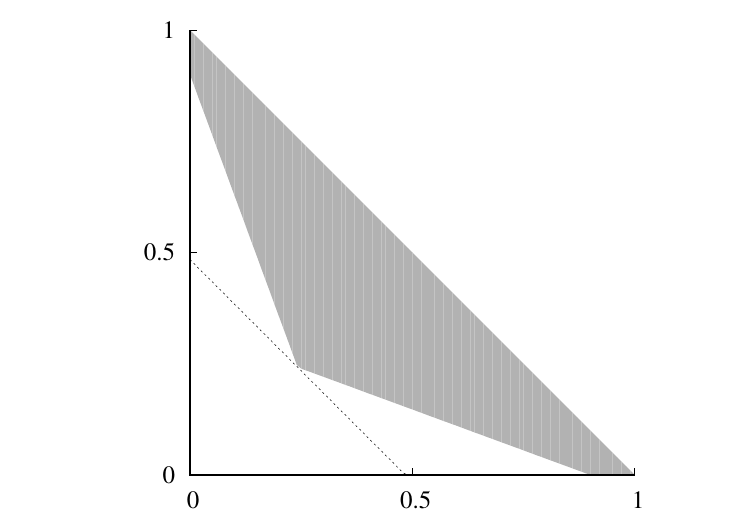}
		\put(-150,135){\footnotesize$\PFA$}
		\put(-35,-3){\footnotesize$\PMD$}
		\put(-190,109){\footnotesize$(0,1-\delta)$}
		\put(-190,44){\footnotesize$(0,\frac{2(1-\delta)}{1+e^{\varepsilon}})$}
		\put(-110,54){\footnotesize$(\frac{(1-\delta)}{1+e^{\varepsilon}},\frac{(1-\delta)}{1+e^{\varepsilon}})$}
		\put(-114,42){$\swarrow$}
		\put(-153,55){$\nearrow$}
		\put(-153,108){$\rightarrow$}
	\end{center}
	\caption{Privacy region for ($\varepsilon,\delta$)-differential privacy. Dotted line represents the solution of
	a maximization problem \eqref{eq:max}.
	For simplicity, we only show the privacy region below the line
	$\PFA+\PMD\leq 1$,
	since the whole region is symmetric w.r.t. the line $\PFA+\PMD = 1$.}
	\label{fig:region1}
	\end{figure}
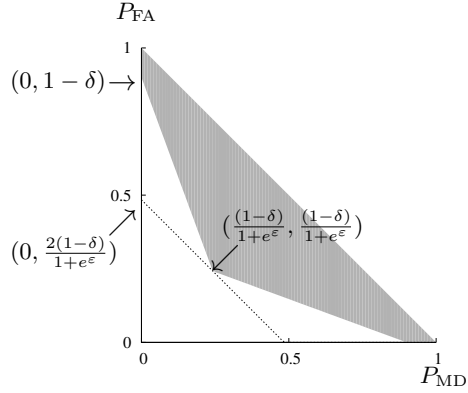
	\begin{eqnarray}
		\cR(\varepsilon,\delta) &\equiv& \big\{(\PMD,\PFA) \,\big|\,
		\PFA + e^\varepsilon \PMD \geq 1-\delta\;, \text{ and }  \;\;
		e^{\varepsilon}\PFA + \PMD \geq 1-\delta \} \;. \label{eq:defregion1}
	\end{eqnarray}	
	Similarly, we define the {\em privacy region} of a database access mechanism $M$
	with respect to two neighboring databases $D$ and $D'$ as
	\begin{eqnarray}
		\cR(M,D,D') &\equiv& \conv\Big( \big\{(\PMD(D,D',M,S),\PFA(D,D',M,S)) \,\big|\,
		\text{ for all } S\subseteq\cX \big\} \Big) \;, \label{eq:defregion2}
	\end{eqnarray}
	where $\conv(\cdot)$ is the convex hull of a set.
	Operationally, by taking the convex hull, the region includes
	the pairs of false alarm and missed detection probabilities
	achieved by soft decisions that might
	use internal randomness in the hypothesis testing.
	Precisely, let $\gamma:\cX\to\{H_0,H_1\}$ be any decision rule where
	we allow probabilistic decisions.
	For example, if the output is in a set $S_1$ we can accept the null hypothesis
	with a certain probability $p_1$, and for another set $S_2$ accept with probability $p_2$.
	In full generality, a decision rule $\gamma$ can be
	fully described by a partition $\{S_i\}$ of the output space $\cX$,
	and corresponding accept probabilities $\{p_i\}$.
	The probabilities of false alarm and missed detection for a decision rule $\gamma$
	is defined as
	$\PFA(D_0,D_1,M,\gamma) \equiv \prob(\gamma(M(D_0))=H_1 )$ and
	$\PMD(D_0,D_1,M,\gamma) \equiv \prob(\gamma(M(D_1))=H_0 )$.
	\begin{remark}
		\label{rem:convexhull}
		For all neighboring databases $D$ and $D'$, and
		a database access mechanism $M$,
		the pair of a false alarm and a missed detection probabilities achieved
		by any decision rule $\gamma$ is included in the privacy region:
		\begin{eqnarray*}
			(\PMD(D,D',M,\gamma),\PFA(D,D',M,\gamma)) &\in& \cR(M,D,D')\;,
		\end{eqnarray*}
		for all decision rule $\gamma$.
	\end{remark}
	Let $D\sim D'$ denote that the two databases are neighbors.
	The union over all neighboring databases define the {\em privacy region of the mechanism}.
	\begin{eqnarray*}
		\cR(M) &\equiv& \bigcup_{D\sim D'} \cR(M,D,D') \;.
	\end{eqnarray*}
	The following corollary, which follows immediately from
	Theorem \ref{thm:hypo}, gives a necessary and sufficient condition on the privacy region
	for $(\varepsilon,\delta)$-differential privacy.
\begin{coro}
	A mechanism $M$ is $(\varepsilon,\delta)$-differentially private if and only if
	$\cR(M) \subseteq \cR(\varepsilon,\delta)$.
	\label{coro:hypo}
\end{coro}
To illustrate the strengths of the graphical representation of differential privacy,
we provide simpler proofs for some well-known results in differential privacy in Appendix \ref{sec:simpleproof}.

Consider two database access mechanisms $M(\cdot)$ and $M'(\cdot)$.
Let $X$ and $Y$ denote the random outputs of mechanisms $M$ and $M'$
respectively.
We say $M$ {\em dominates} $M'$ if $M'(D)$ is conditionally independent of the database $D$
conditioned on the outcome of $M(D)$.
In other words, the database $D$, $X=M(D)$ and $Y=M'(D)$
form the following Markov chain: $D$--$X$--$Y$.
	
\begin{thm}[Data processing inequality for differential privacy]
	\label{thm:dpi}
	If a mechanism $M$ dominates a mechanism $M'$, then
	for all pairs of neighboring databases $D_1$ and $D_2$,
	\begin{eqnarray*}
		\cR(M',D_1,D_2) &\subseteq& \cR(M,D_1,D_2)\;.
	\end{eqnarray*}
\end{thm}
We provide a proof in Section \ref{sec:proofdpi}.
Wasserman and Zhu  \cite[Lemma 2.6]{WZ10} have proved
that, for a special case when $M$ is $(\varepsilon,0$)-differentially private,
$M'$ is also $(\varepsilon,0$)-differentially private,
which is a corollary of the above theorem.
Perhaps surprisingly, the converse is also true.

\begin{thm}[{\cite[Corollary of Theorem 10]{Bla53}}]
	\label{thm:converse}
	Fix a pair of neighboring databases $D_1$ and $D_2$
	and let $X$ and $Y$ denote the random outputs of mechanisms $M$ and $M'$,
	respectively.
	If  $M$ and $M'$ satisfy
	\begin{eqnarray*}
		\cR(M',D_1,D_2) &\subseteq& \cR(M,D_1,D_2)\;,
	\end{eqnarray*}
	then there exists a coupling of the random outputs $X$ and $Y$ such that
	they form a Markov chain $D$--$X$--$Y$ where $D\in\{D_1,D_2\}$.
\end{thm}
%
When the privacy region of $M'$ is included in $M$, then there exists
a stochastic transformation $T$
that operates on $X$ and produce a random output that has the same marginal distribution as $Y$ conditioned on the database $D$.
We can consider this mechanism $T$ as a  privatization mechanism that
takes a (privatized) output $X$ and
provides even further privatization.
The above theorem was proved in \cite[Corollary of Theorem 10]{Bla53} in the context of
comparing two experiments, where
a statistical {\em experiment} denotes a mechanism in the context of  differential privacy.

%
%

\section{Composition of Differentially Private Mechanisms}

In this section, we address how differential privacy guarantees compose:
when accessing databases multiple times via differentially private mechanisms,
each of which having its own privacy guarantees,
how much privacy is still guaranteed on the union of those outputs?
To formally define composition, we consider the following scenario known as the `composition experiment',
proposed in \cite{DRV10}.

A composition experiment takes as input a parameter $b\in\{0,1\}$,
and an adversary $\cA$.
From the hypothesis testing perspective proposed in the previous section,
$b$ can be interpreted as the hypothesis: null hypothesis for $b=0$ and
alternative hypothesis for $b=1$.
At each time $i$, a database $D^{i,b}$ is accessed depending on $b$.
For example, one includes a particular individual and another does not.
An adversary $\cA$ is trying to break privacy
(and figure out whether the particular individual is in the database or not) by
testing the hypotheses on the output of
$k$ sequential access to those databases via differentially private mechanisms.
In full generality, we allow the adversary to have full control over which pair of databases to access,
which query to ask,
and which mechanism to be used at each repeated access.
Further, the adversary is free to make these choices adaptively based on the previous outcomes.
The only restrictions are
the differentially private mechanisms belong to a family $\cM$ (e.g., the family of all
$(\varepsilon,\delta)$-differentially private mechanisms),
the internal randomness of the mechanisms are independent at each repeated access,
and that the hypothesis $b$ is not known to the adversary.

\begin{center}
\begin{tabular}{ll}
\hline
\vspace{-.35cm}\\
\multicolumn{2}{l}{ \compose }\\
\hline
\vspace{-.35cm}\\
\multicolumn{2}{l}{{\bf Input:} $\cA$, $\cM$, $k$, $b$} \\
\multicolumn{2}{l}{{\bf Output:} $V^b$}\\
 & {\bf for } $i=1$ to $k$ {\bf do}\\
   & \hspace{0.6cm}  $\cA$ requests $(D^{i,0},D^{i,1},q_i,M_i)$ for some $M_i\in\cM$; \\
   & \hspace{0.6cm}  $\cA$ receives $y_i = M_i(D^{i,b},q_i)$; \\
  & {\bf end for } \\
 & Output the view of the adversary  $V^b = (R^b,Y_1^b,\ldots,Y_k^b)$.\\
\hline
\end{tabular}
\end{center}

The outcome of this $k$-fold composition experiment is
the {\em view of the adversary} $\cA$:
$V^b \equiv (R,Y_1^b,\ldots,Y_k^b)$,
which is the sequence of random outcomes $Y_1^b,\ldots,Y_k^b$,
and the outcome $R$ of any internal randomness of $\cA$.

\subsection{Optimal privacy region under composition}
In terms of testing whether a particular individual is
in the database $(b=0)$ or not $(b=1$),
we want to characterize how much privacy degrades
after a $k$-fold composition experiment.
It is known that the privacy degrades under composition
by at most  the `sum' of the differential privacy parameters
of each access.
\begin{thm}[{\cite{DMNS06,DKM06,DL09,DRV10}}]
	\label{thm:ke}
	For any $\varepsilon>0$ and $\delta\in[0,1]$,
	the class of $(\varepsilon,\delta)$-differentially private mechanisms
	satisfy $(k\varepsilon,k\delta)$-differential privacy under $k$-fold adaptive composition.
\end{thm}
In general, one can show that if $M_i$ is $(\varepsilon_i,\delta_i$)-differentially private,
then the composition satisfies
$(\sum_{i\in[k]}\varepsilon_i,\sum_{i\in[k]}\delta_i)$-differential privacy.
If we do not allow any slack in the $\delta$, this bound cannot be tightened.
Precisely, there are examples of mechanisms which under $k$-fold composition
violate $(\varepsilon,\sum_{i\in[k]}\delta_i)$-differential privacy for
any $\varepsilon<\sum_{i\in[k]}\varepsilon_i$.
We can prove this by providing a set $S$ such that
the privacy condition is met with equality:
$\prob(V^0 \in S) = e^{\sum_{i\in[k]}\varepsilon_i} \prob(V^1\in S) + \sum_{i\in[k]}\delta_i$.
However, if we allow for a slightly larger value of $\delta$,
then Dwork et al. showed in \cite{DRV10} that
one can gain a significantly
higher privacy guarantee in terms of $\varepsilon$.
\begin{thm}[{\cite[Theorem III.3]{DRV10}}]
	\label{thm:boosting}
	For any $\varepsilon>0$, $\delta\in[0,1]$, and $\tdelta\in(0,1]$,
	the class of  $(\varepsilon,\delta)$-differentially private mechanisms
	satisfies $(\tepsilon_\tdelta,k\delta+\tdelta)$-differential privacy under $k$-fold adaptive composition, for
	\begin{eqnarray}
		\tepsilon_\tdelta &=& k\varepsilon(e^\varepsilon-1)+\varepsilon\sqrt{2k\log(1/\tdelta)}.
		\label{eq:boosting}
	\end{eqnarray} 	
\end{thm}
By allowing a slack of $\tdelta>0$, one can get a  higher privacy of
 $\tepsilon_\tdelta=O(k\varepsilon^2 + \sqrt{k\varepsilon^2})$, which is
 significantly smaller than $k\varepsilon$.
 This is the best known guarantee so far,
 and has been used whenever one requires a
 privacy guarantee under composition (e.g. \cite{DRV10,BBDS12,HR13}).
 However, the important question of
 optimality has remained open.
 Namely, is there a composition of mechanisms where the above
 privacy guarantee is tight? In other words, is it possible to get a tighter bound on
 differential privacy under composition?

We give a complete answer to this fundamental question in the following theorems.
We prove a tighter bound on the privacy under composition.
Further,
we also prove the achievability of the privacy guarantee:
we provide a set of mechanisms
such that the privacy region under
$k$-fold composition is exactly the region defined by the
conditions in \eqref{eq:composition}.
Hence, this bound on the privacy region is tight and cannot be improved upon.

\begin{thm}
	\label{thm:composition}
	For any $\varepsilon \geq 0$ and $\delta\in[0,1]$,
	the class of $(\varepsilon,\delta)$-differentially private mechanisms
	satisfies
	\begin{eqnarray}
		\big(\, (k-2i)\varepsilon \,,\,1-(1-\delta)^k(1-\delta_i)\, \big)\text{-differential privacy} \; \label{eq:composition}
	\end{eqnarray}
	under $k$-fold adaptive composition,
	for all $i=\{0,1,\ldots,\lfloor k/2 \rfloor\}$,
	where
	\begin{eqnarray}
		\label{eq:defed}
		\delta_i &=& \frac{\sum_{\ell=0}^{i-1} { k\choose \ell} \big(e^{(k-\ell)\varepsilon}-e^{(k-2i+\ell)\varepsilon}\big) }{(1+e^\varepsilon)^k} \;.
	\end{eqnarray}
\end{thm}
Hence, the privacy region of $k$-fold composition is
an intersection of $k$ regions, each of which is $((k-2i)\varepsilon, 1-(1-\delta)^k(1-\delta_i) )$-differentially private:
$\cR(\{(k-2i)\varepsilon, 1-(1-\delta)^k(1-\delta_i) \}_{i\in[k/2]})\, \equiv \bigcap_{i=0}^{\lfloor \frac{k}{2}\rfloor} \cR((k-2i)\varepsilon, 1-(1-\delta)^k(1-\delta_i) )$.
We give a proof in Section \ref{sec:compositionproof}
where we give an explicit mechanism that achieves this region under composition.
Hence, this bound on the privacy region is tight, and
gives the exact description of how much privacy can degrade under $k$-fold adaptive composition.
This settles the question left open in \cite{DMNS06,DKM06,DL09,DRV10}
by providing, for the first time, the fundamental limit of composition, and
proving a matching mechanism with the worst-case privacy degradation.

\begin{figure}[t]
\begin{center}
	\includegraphics[width=.4\textwidth]{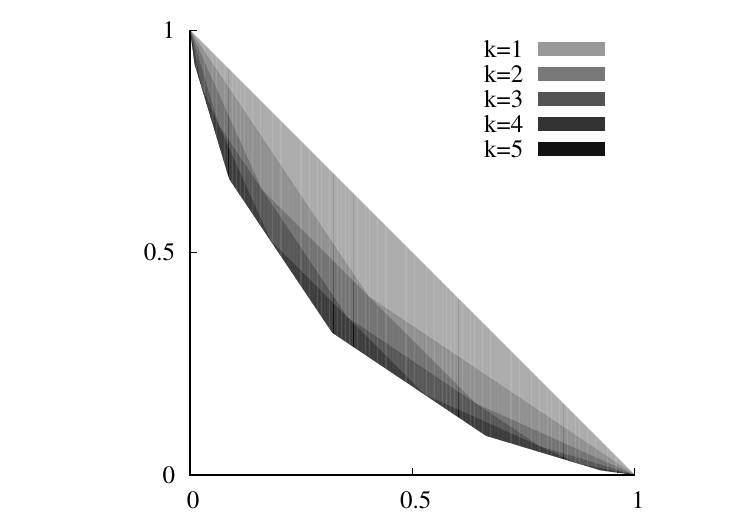}
		\put(-165,115){\footnotesize$\PFA$}
		\put(-35,-3){\footnotesize$\PMD$}
	\includegraphics[width=.4\textwidth]{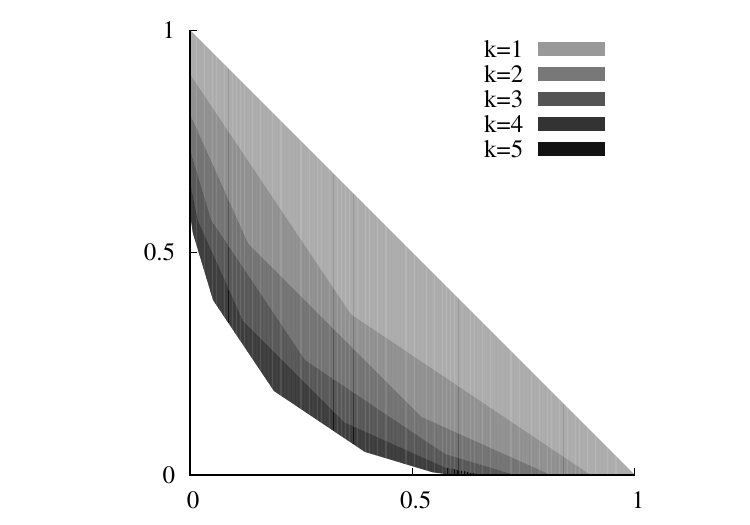}
		\put(-165,115){\footnotesize$\PFA$}
		\put(-35,-3){\footnotesize$\PMD$}
\end{center}
\caption{Privacy region $\cR(\{(k-2i)\varepsilon,\delta_i\})$
for the class of ($\varepsilon,0$)-differentially private mechanisms (left) and
($\varepsilon,\delta$)-differentially private mechanisms (right) under $k$-fold adaptive composition.}
\label{fig:compose}
\end{figure}

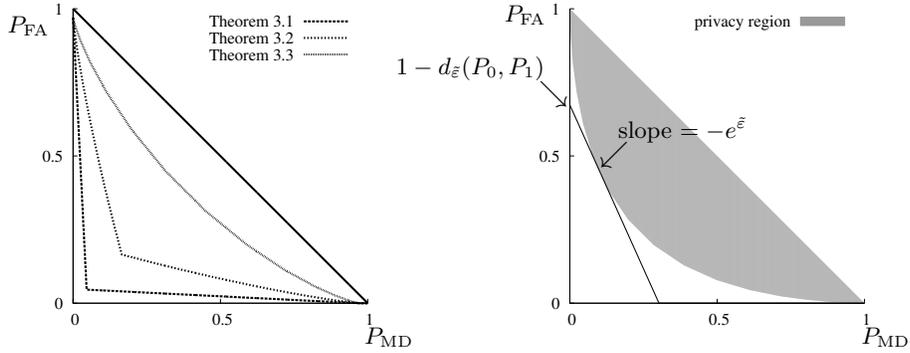
\begin{figure}[t]
	\begin{center}
	\includegraphics[width=.4\textwidth]{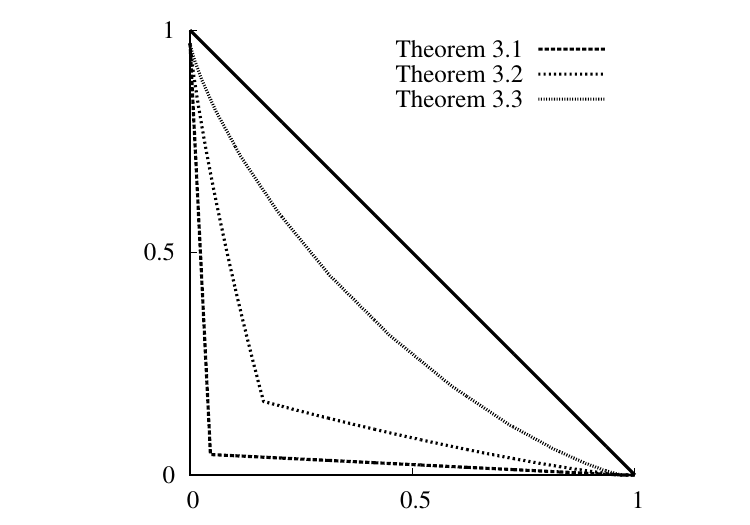}
		\put(-165,115){\footnotesize$\PFA$}
		\put(-30,-3){\footnotesize$\PMD$}
	\includegraphics[width=.4\textwidth]{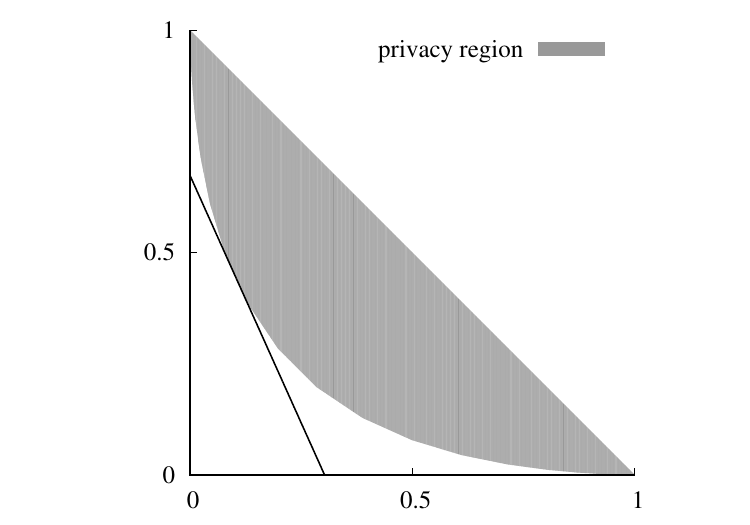}
		\put(-165,119){\footnotesize$\PFA$}
		\put(-30,-3){\footnotesize$\PMD$}
		\put(-206,99){\footnotesize$1-d_{\tepsilon}(P_0,P_1)$}
		\put(-152,89){$\searrow$}
		\put(-122,75){\footnotesize slope $= -e^\tepsilon$}
		\put(-129,66){$\swarrow$}
	\end{center}
	\caption{Theorem \ref{thm:composition} provides the tightest bound (left).
	Given a mechanism $M$,  the privacy region can be completely described by its boundary,
	which is represented by a set of tangent lines of the form
	$\PFA = -e^\tepsilon \PMD + 1-d_\tepsilon(P_0,P_1)$ (right). }
	\label{fig:tangent}
\end{figure}

To prove the optimality of our main result in Theorem \ref{thm:composition},
namely that it is impossible to have a privacy worse than \eqref{eq:composition},
we rely on the operational interpretation of the privacy as hypothesis testing.
To this end, we use the new analysis tools (Theorem \ref{thm:dpi} and Theorem \ref{thm:converse})
provided in the previous section.
Figure \ref{fig:compose} illustrates how much the privacy region of Theorem \ref{thm:composition}
degrades
as we increase the number of composition $k$.
Figure \ref{fig:tangent} provides a comparison of the
three privacy guarantees in Theorems \ref{thm:ke},
\ref{thm:boosting} and \ref{thm:composition}
for $30$-fold composition of $(0.1,0.001)$-differentially private mechanisms.
Smaller region gives a tighter bound, since it guarantees the higher privacy.

\subsection{Simplified privacy region under composition }
\label{sec:simplified}

In many applications of the composition theorems,
a closed form expression of the composition privacy guarantee is required.
The privacy guarantee in \eqref{eq:composition} is tight, but can be difficult to evaluate.
The next theorem provides a simpler form expression which is an outer bound of the
exact region described in \eqref{eq:composition}.
Comparing to \eqref{eq:boosting}, the privacy guarantee is significantly improved
from $\tepsilon_\tdelta = O\Big(k\varepsilon^2+\sqrt{k\varepsilon^2\log(1/\tdelta)}\Big)$ to
$\tepsilon_\tdelta = O\Big(k\varepsilon^2 + \min \big\{\sqrt{k\varepsilon^2\log(1 /\tdelta)}, \varepsilon \log(\varepsilon/\tdelta) \big\} \Big)$, especially when composing a large number $k$ of interactive queries.
Further, the $\delta$-approximate differential privacy degradation of $(1-(1-\delta)^k(1-\tdelta))$ is also strictly smaller than
the previous $(k\delta+\tdelta)$.
We discuss the significance of this improvement  in the next section
using examples from existing differential privacy literature.

\begin{thm}
	\label{thm:closedform}
	For any $\varepsilon>0$, $\delta\in[0,1]$, and $\tdelta\in[0,1]$,
	the class of $(\varepsilon,\delta)$-differentially private mechanisms
	satisfies $\big(\tepsilon_\tdelta,1-(1-\delta)^k(1-\tdelta)\big)$-differential
	privacy under $k$-fold adaptive composition, for
	\begin{eqnarray}
		\tepsilon_\tdelta &=& \min \left\{\, k\varepsilon \,\,,\,
		\frac{(e^\varepsilon-1)\varepsilon k}{e^\varepsilon+1} + \varepsilon\sqrt{2k\,\log\Big(e+\frac{\sqrt{k\varepsilon^2}}{\tdelta}\,\Big)}
		\,\,,\,\,
			  \frac{(e^\varepsilon-1)\varepsilon k}{e^\varepsilon+1} + \varepsilon\sqrt{2k\,\log\Big(\frac{1}{\tdelta}\,\Big)} \, \right\} \;.
			  \label{eq:closedform}
	\end{eqnarray} 	
\end{thm}
In the high privacy regime, where $\varepsilon\leq 0.9$,
this bound can be further simplified as
\begin{eqnarray*}
	\tepsilon_\tdelta \;\;  \leq \;\; \min \Big\{\, k\varepsilon, k\varepsilon^2 + \varepsilon \sqrt{2k\log\big(e+(\sqrt{k\varepsilon^2}/\tdelta \,)\,\big)},k\varepsilon^2+\varepsilon\sqrt{2k\log(1/\tdelta)}\,\Big\}\;.
\end{eqnarray*}
A proof  is provided in  Section \ref{sec:closedform}.
This privacy guarantee improves over the existing result of
Theorem \ref{thm:boosting} when $\tdelta=\Theta(\sqrt{k\varepsilon^2})$.
Typical regime of interest is the high-privacy regime for composition privacy guarantee,
i.e. when $\sqrt{k\varepsilon^2} \ll 1$. The above theorem suggests that
we only need the extra slack of approximate privacy $\tdelta$ of order $\sqrt{k\varepsilon^2}$.

\subsection{Composition Theorem for Heterogeneous Mechanisms}

We considered homogeneous mechanisms, where all mechanisms are
$(\varepsilon,\delta)$-differentially private. Our analysis readily extends to
heterogeneous mechanisms, where the $\ell$-th query satisfies
($\varepsilon_\ell,\delta_\ell$)-differential privacy (we refer  to such mechanisms as
$(\varepsilon_\ell,\delta_\ell)$-differentially private mechanisms).
\begin{thm}
	\label{thm:hetero}
	For any $\varepsilon_\ell>0$, $\delta_\ell\in[0,1]$ for $\ell\in\{1,.\ldots,k\}$, and $\tdelta\in[0,1]$,
	the class of $(\varepsilon_\ell,\delta_\ell)$-differentially private mechanisms
	satisfy $\big(\tepsilon_\tdelta,1-(1-\tdelta)\prod_{\ell=1}^k(1-\delta_\ell) \big)$-differential
	privacy under $k$-fold adaptive composition, for $\tepsilon_\tdelta = $
	\begin{eqnarray}
		  \min \left\{\, \sum_{\ell=1}^k \varepsilon_\ell \,\,,\,
			\,
			 \sum_{\ell=1}^k \frac{(e^{\varepsilon_\ell}-1)\varepsilon_\ell }{e^{\varepsilon_\ell}+1} + \sqrt{\sum_{\ell=1}^k 2\,\varepsilon_\ell^2 \,\log\Big(e+\frac{\sqrt{\sum_{\ell=1}^k  \varepsilon_\ell^2}}{\tdelta}\,\Big)}
			 \,\,,\, \,
			  \sum_{\ell=1}^k \frac{(e^{\varepsilon_\ell}-1)\varepsilon_\ell }{e^{\varepsilon_\ell}+1} + \sqrt{\sum_{\ell=1}^k 2\,\varepsilon_\ell^2 \,\log\Big(\frac{1}{\tdelta}\,\Big)} \, \right\} \;.
			  \label{eq:hetero}
	\end{eqnarray} 	
\end{thm}
This tells us that the $\varepsilon_\ell$'s {\em sum up under composition}:
whenever we have $k\varepsilon$ or $k\varepsilon^2$ in
\eqref{eq:closedform} we can replace it by the summation to get the general result for heterogeneous case.

%
%
\section{Applications of the Optimal Composition Theorem}
\label{sec:application}

When analyzing a complex mechanism with multiple sub-mechanisms each
with ($\varepsilon_0,\delta_0$)-differential privacy guarantee,
we can apply the composition theorem (Theorem \ref{thm:composition} and Theorem \ref{thm:closedform}).
To ensure overall $(\varepsilon,\delta)$-differential privacy for the whole complex mechanism,
one chooses $\varepsilon_0 = \varepsilon/(2\sqrt{k \log(e+\varepsilon/\delta)})  $
and $\delta_0 = \delta/2k$, when there are $k$ sub-mechanisms.
The existing composition theorem guarantees the desired overall privacy.
Then, the {\em utility} of the complex mechanism is
calculated for the choice of $\varepsilon_0$ and $\delta_0$.

Following this recipe, we first provide a sufficient condition on the variance of noise adding mechanisms.
This analysis shows that one requires  smaller variance than what is previously believed,
in the regime where $\varepsilon=\Theta(\delta)$. Further, we show that a variety of known mechanisms
achieve the desired privacy under composition with the {\em same} level of variance.
Applying this analysis to  known mechanisms for cut queries of a graph,
we show that again in the regime where  $\varepsilon=\Theta(\delta)$,
one can achieve the desired privacy under composition with improved utility.

For count queries with sensitivity one, the geometric noise adding mechanism is known to be universally optimal in a general cost minimization framework (Bayesian setting in \cite{GRS12} and worst-case setting in \cite{GV12}). Here we provide
 a new interpretation of the geometric noise adding mechanism as an optimal mechanism under {\em composition} for counting queries. In the course of proving Theorem \ref{thm:composition}, we show that
a family of mechanisms are optimal under composition, in the sense that they achieve the
largest privacy region among $k$-fold compositions of any $(\varepsilon_i,\delta_i)$ differentially private mechanisms.
Larger region under composition implies that one can achieve smaller error rates,
while ensuring the same level of privacy at each step of the composition.
In this section, we show that the geometric mechanism is one of such mechanisms, thus providing the new interpretation to the optimality of the geometric mechanisms.

%
%
\subsection{Variance of noise adding mechanisms under composition}
\label{sec:variance}

In this section, we consider real-valued queries $q: \cD \to \reals$.
The {\em sensitivity} of a real-valued query is defined as the maximum absolute difference of the
output between two neighboring databases:
\begin{eqnarray*}
	\Delta &\equiv& \max_{D\sim D'} |q(D)-q(D')|\;,
\end{eqnarray*}
where $\sim$ indicates that the pair of databases are neighbors.
A common approach to privatize such a query output is to add noise to it,
and the variance of the noise grows with sensitivity of the query and the desired level of privacy.
A popular choice of the noise is Gaussian.
It is previously known that
it is sufficient to add Gaussian noise with variance $O(k\Delta^2\log(1/\delta)/\varepsilon^2)$
to each query output
in order to ensure $(\varepsilon,\delta)$-differential privacy under $k$-fold composition.
We improve the analysis of Gaussians under composition,
and show that for a certain regime where $\varepsilon=\Theta(\delta)$,
the sufficient condition can be improved by a log factor.

When composing real-valued queries,
the {\em Gaussian mechanism} is a popular choice \cite{DN03,DN04,BDMN05,BBDS12,HR13}.
However,
we show that there is nothing special about Gaussian mechanisms for composition.
We prove that the {\em Laplacian mechanism}
or the {\em staircase mechanism} introduced in \cite{GV12} can achieve the same level of privacy
under composition with the same variance.

We can use Theorem \ref{thm:closedform}
to find how much noise we need to add to each query output,
in order to ensure $(\varepsilon,\delta)$-differential privacy under $k$-fold composition.
We know that if each query output is $(\varepsilon_0,\delta_0)$-differentially private,
then the composed outputs satisfy
$(k\varepsilon_0^2 + \sqrt{2k\varepsilon_0^2\log(e+\sqrt{k\varepsilon_0^2}/\tdelta)} ,k\delta_0+\tdelta)$-differential privacy
assuming $\varepsilon_0\leq0.9$.
With the choice of $\delta_0=\delta/2k$, $\tdelta=\delta/\sqrt{2}$,
and $\varepsilon_0^2 = \varepsilon^2/4k\log(e+(\varepsilon/\delta)) $,
this ensures that the target privacy of $(\varepsilon,\delta)$ is satisfied under $k$-fold composition as described in the
following corollary.
\begin{coro}
	\label{coro:closedform}
	For any $\varepsilon\in(0,0.9]$ and $\delta\in(0,1]$,
	if the database access mechanism satisfies
	$(\sqrt{\varepsilon^2/4k\log(e+(\varepsilon/\delta)) }, \delta/2k )$-differential privacy on each query output,
	then it satisfies $(\varepsilon,\delta)$-differential privacy under $k$-fold composition.
\end{coro}

One of the most popular noise adding mechanisms is the {\em Laplacian mechanism},
which adds Laplacian noise to real-valued query outputs.
When the sensitivity is $\Delta$,
one can achieve $(\varepsilon_0,0)$-differential privacy with
 the choice of the distribution
 ${\rm Lap}(\varepsilon_0/\Delta) = (\varepsilon_0/2\Delta)e^{-\varepsilon_0|x|/\Delta}$.
 The resulting variance of the noise is $2\Delta^2/\varepsilon_0^2$.
The above corollary implies a certain sufficient condition on the variance of
the Laplacian mechanism to ensure
 privacy under composition.
\begin{coro}
	\label{coro:Laplacian}
	For real-valued queries with sensitivity $\Delta>0$,
	the mechanism that adds
	Laplacian noise with variance $\big(8k\Delta^2\log\big(e+(\varepsilon/\delta)\big)/\varepsilon^2\big)$
	satisfies $(\varepsilon,\delta)$-differential privacy under $k$-fold adaptive composition
	for any $\varepsilon\in(0,0.9]$ and $\delta\in(0,1]$.
\end{coro}

In terms of variance-privacy trade-off for real-valued queries,
the optimal noise-adding mechanism known as
the {\em staircase mechanism} was introduced in \cite{GV12}.
The probability density function of this noise is piecewise constant, and
the probability density on the pieces decay geometrically.
It is shown in \cite{GV13} that that with
variance of $O(\min \{ 1/\varepsilon^2, 1/\delta^2\} )$,
the staircase mechanism achieved $(\varepsilon,\delta)$-differential privacy.
Corollary \ref{coro:closedform} implies that
with variance $O\big(k\Delta^2 \log(e+\varepsilon/\delta)/\varepsilon^2\big)$,
the staircase mechanism satisfies $(\varepsilon,\delta)$-differential
privacy under $k$-fold composition.

Another popular mechanism known as the {\em Gaussian mechanism} privatizes each query output
by adding a Gaussian noise with variance $\sigma^2$.
It is not difficult to show that when the sensitivity of the query is $\Delta$,
with a choice of $\sigma^2\geq 2\Delta^2 \log(2/\delta_0)/\varepsilon_0^2$,
the Gaussian  mechanism satisfies $(\varepsilon_0,\delta_0)$-differential privacy (e.g. \cite{DKM06}).
The above corollary implies that
the Gaussian mechanism with variance $O(k\Delta^2\log(1/\delta)\log(e+(\varepsilon/\delta))/\varepsilon^2)$
ensures $(\varepsilon,\delta)$-differential  privacy under $k$-fold composition.
However, we can get a tighter sufficient condition by directly analyzing
how Gaussian mechanisms compose, and the proof is provided in Appendix \ref{sec:Gaussianproof}.
\begin{thm}
	\label{thm:Gaussian}
	For real-valued queries with sensitivity $\Delta>0$,
	the mechanism that adds
	Gaussian noise with variance $\big(8k\Delta^2\log\big(e+(\varepsilon/\delta)\big)/\varepsilon^2\big)$
	satisfies $(\varepsilon,\delta)$-differential privacy under $k$-fold adaptive composition
	for any $\varepsilon>0$ and $\delta\in(0,1]$.
\end{thm}
It is previously known that it is sufficient to add i.i.d. Gaussian noise with
variance $O(k \Delta^2 \log(1/\delta) /\varepsilon^2)$ to ensure
$(\varepsilon,\delta)$-differential privacy under $k$-fold composition
(e.g. \cite[Theorem 2.7]{HT10}).
The above theorem shows that when $\delta=\Theta(\varepsilon)$,
one can achieve the same privacy with smaller variance by a
factor of $\log(1/\delta)$.

%
%
\subsection{Geometric noise adding mechanism under composition}
\label{sec:geometric}

In this section, we consider integer valued queries $q:\cD\to\Z$ with
sensitivity one, also called {\em counting queries}.
Such queries are common in practice,
e.g. ``How many individuals have income less than \$100,000?''.
Presence of absence of an individual record changes the output at most by one.
Counting query is a well-studied topic in differential privacy \cite{DN03,DN04,BDMN05,BLR13} and
they provide a primitive for constructing more complex queries \cite{BDMN05}.

The {\em geometric} noise adding mechanism is a discrete variant of the popular Laplacian mechanism.
For integer-valued queries with sensitivity one,
the mechanism adds a noise distributed according to a
double-sided geometric distribution whose probability density function is
$p(k)=\big( (e^\varepsilon-1)/(e^\varepsilon+1) \big) e^{-\varepsilon|k|}$.
This mechanism is known to be universally optimal in a general cost minimization framework (Bayesian setting in \cite{GRS12} and worst-case setting in \cite{GV12}).
In this section, we show that the geometric noise adding mechanism 
achieves the fundamental limit on the  privacy region under composition.

Consider the composition experiment for counting queries.
For a pair of neighboring databases $D_0$ and $D_1$,
some of the query outputs differ by one, since sensitivity is one,
and for other queries the output might be the same.
Let $k$ denote the number of queries whose output differs with respect to $D_0$ and $D_1$.
Then, we show in Section \ref{sec:geometricproof} that
the privacy region achieved by geometric mechanism,
that adds geometric noise for each integer-valued query output,
is exactly described by the optimal composition theorem of \eqref{eq:composition}.
Further, since this is the largest privacy region under composition
for the pair of database $D_0$ and $D_1$ that differ in $k$ queries,
no other mechanism can achieve a larger privacy region.
Since the geometric mechanism does not depend on the particular choice of pairs of databases $D_0$ and $D_1$,
nor does it depend on the specific query being asked,
the mechanism achieves the exact composed privacy region universally for every pair of neighboring databases simultaneously.

Among the mechanisms guaranteeing the same level of privacy,
one with larger privacy region under composition is considered better, in terms of allowing for
smaller false alarm and missed detection rate in hypothesis testing whether the database contains a particular entry or not.
In this sense, larger privacy degradation under composition has more utility.
The geometric mechanism has the largest possible privacy degradation under composition, stated formally below; the proof is deferred to Appendix~\ref{sec:geometricproof}.

\begin{thm}
 	Under the $k$-fold composition experiment of counting queries,
	the geometric mechanism achieves the largest privacy region among all $(\varepsilon,0)$-differentially private mechanisms,
	universally for every pair of neighboring databases simultaneously.
	\label{thm:geometric}
\end{thm}

%
%
\section{Applications of the Operational Interpretation to Private Multi-Party Computation}
\label{sec:multiparty}
In this section, we showcase the power of the operational interpretation of differential privacy in the differentially private multi-party computation (MPC) setting \cite{BNO08,DKMMN,MMPT,GMPS}. 
We study the following problem of secure multi-party differential privacy:
each party possesses a single bit of information; the information
bits are statistically independent. Each party is interested in computing a
function, which could differ from party to party, and there could be a
central observer (observing the entire transcript of the interactive
communication protocol) interested in computing a separate function.
The interactive communication is achieved via a broadcast channel that all parties and central observer can hear. 
It is useful to distinguish between two types of communication protocols: {\em interactive} and {\em non-interactive}.
We say a communication protocol is non-interactive if a message broadcasted by one party does not depend on
the messages broadcasted by any other parties. In contrast, interactive protocols allows the messages at any stage of the communication to
depend on all the previous messages.

Our main result is the exact optimality of a simple non-interactive protocol in terms of maximizing accuracy for given privacy levels: each party randomizes (sufficiently) and publishes its own bit.  Each party and the central observer then separately compute their respective decision functions to maximize the appropriate notion of their accuracy measure. The optimality is general: it holds for all types of functions, heterogeneous privacy conditions on  the parties, all types of cost metrics, and both average and worst-case (over the inputs) measures of accuracy. Finally, the optimality result is {\em simultaneous}, in terms of maximizing accuracy at each of the parties and the central observer. Each party only needs to know its own desired level of privacy, its own function to be computed, and its measure of accuracy. Optimal data release and optimal decision making are naturally separated.



The proof of this result critically relies on the operational interpretation of differential privacy. In this multi-party and local privacy setting, we show that the randomized response still dominates any other $(\varepsilon,\delta)$-differentially private mechanisms. Given this, any other mechanism, interactive or not, can be simulated at the receiver. 
This powerful technique bypasses the previous results on the same setting, 
where weaker results were proved with heavier proof techniques. 
In \cite{GMPS}, optimal mechanisms are proposed for only two-party  computation and only for 
AND and XOR functions. 
In \cite{KOV15},  only $(\varepsilon,0)$-differential privacy is addressed, and the proof techniques developed in \cite{KOV15} cannot be generalized to the more general $(\varepsilon,\delta)$-differential privacy setting. 

\subsection{Problem Statement}
Consider the setting where there are $k$ parties, each with its own private binary  data
$x_i\in\{0,1\}$ generated independently.
The independence assumption here is necessary
because without it each party can learn something about others,
which violates differential privacy, even without revealing any information. Differential privacy implicitly imposes independence in a multi-party setting.
The goal of each party $i\in[k]$ is to compute
an arbitrary function $f_i:\{0,1\}^k\to \cY$ of interest
by interactively broadcasting messages.
There might be a central observer who listens to all the messages being broadcasted,
and wants to compute another arbitrary function $f_0:\{0,1\}\to\cY$.
The $k$ parties  are honest in the sense that once they agree on what protocol to follow,
every party follows the rules.
At the same time, they can be curious, and
each party needs to ensure that other parties cannot learn its bit
with sufficient confidence.
This is done by imposing local differential privacy constraints. This setting is similar to the one studied in \cite{DJW13,KOV14-1} in the sense that
there are multiple privacy barriers, each one separating
an individual party from the rest of the world.
However, the main difference is that we consider multi-party computation, where
there are multiple functions to be computed, and each node might possess a different function to be computed.

Let $x=[x_1,\ldots,x_k]\in\{0,1\}^k$ denote the vector of $k$ bits,
and $x_{-i}=[x_1,\ldots,x_{i-1},x_{i+1},\ldots,x_k]\in\{0,1\}^{k-1}$ is the vector of bits except
for the $i$-th bit.
The parties agree on an interactive protocol to achieve the goal of multi-party computation.
A `transcript' is the output of the protocol, and
is a random instance of all broadcasted messages until all communication terminates.
The probability that a transcript $\tau$ is broadcasted (via a series of interactive communications)
when the data is $x$
is denoted by
$P_{x,\tau} = \prob(\tau\,|\,x)$ for $x\in\{0,1\}^k$ and for $\tau\in \cT$.
Then, a protocol can be represented as a matrix denoting the probability distribution over
a set of transcripts $\cT$ conditioned on $x$: $P=[P_{x,\tau}] \in[0,1]^{2^k\times|\cT|}$.

In the end, each party makes a decision on what the value of function $f_i$
is, based on its own bit $x_i$ and the transcript $\tau$ that was
broadcasted. A decision rule is a mapping from a transcript $\tau\in\cT$ and
private bit $x_i\in\{0,1\}$ to a decision $y\in\cY$ represented by a function
$\hf_i(\tau,x_i)$. We allow randomized decision rules, in which case
$\hf_i(\tau,x_i)$ can be a random variable. For the central observer, a
decision rule is a function of just the transcript, denoted by a function
$\hf_0(\tau)$.

We consider two notions of accuracy: the average accuracy and the worst-case accuracy.
For the $i$-th party, consider an accuracy measure $\w_i:\cY\times\cY \to \reals$ (or equivalently a negative cost function) such that
$\w_i(f_i(x),\hf_i(\tau,x_i))$ measures the accuracy when the function to be computed is $f_i(x)$ and
the approximation is $\hf_i(\tau,x_i)$.
Then the average  accuracy for this $i$-th party is defined as
\begin{eqnarray}
	\aveacc(P,w_i,f_i,\hf_i) & \equiv & \frac{1}{2^k}\sum_{x\in\{0,1\}^k} \E_{\hf_i,P_{x,\tau} }[\w_i{(f_i(x),\hf_i(\tau,x_i))}] \;,	\label{eq:aveacc}
\end{eqnarray}
where the expectation is taken over the random transcript $\tau$ and any randomness in the decision function $\hf_i$.
For example, if the accuracy measure is an indicator such that $w_i(y,y')=\ind_{(y=y')}$,
then $\aveacc$ measures the average probability of getting the correct function output.
For a given protocol $P$,
it takes $(2^k \,|\cT|)$ operations to compute the optimal decision rule:
\begin{eqnarray}
	f^*_{i,\rm ave}(\tau,x_i) &=& \arg\max_{y\in\cY} \sum_{x_{-i}\in\{0,1\}^{k-1}} P_{x,\tau} \,\w_i(f_i(x),y)\;,\label{eq:optave}
\end{eqnarray}
for each $i\in[k]$. The computational cost of $(2^k \,|\cT|)$ for computing
the optimal decision rule is {\em unavoidable in general}, since that is the
inherent complexity of the problem: describing the distribution of the
transcript requires the same cost. We will show that the optimal protocol
requires a set of transcripts of size $|\cT|=2^k$, and the computational
complexity of the decision rule for a general function is $2^{2k}$. However,
for a fixed protocol, this decision rule needs to be computed only once
before any message is transmitted.
Further, it is also possible to find a closed form solution for the decision rule when
$f$ has a simple structure. One example is the XOR function where
the optimal decision rule is as simple as evaluating the XOR of all the received bits, which requires $O(k)$ operations.
When there are multiple maximizers $y$, we
can choose either one of them arbitrarily, and it follows that there is no gain in randomizing
the decision rule for average accuracy. Similarly, the worst-case accuracy is
defined as
\begin{eqnarray}
	\wcacc(P,\w_i,f_i,\hf_i) & \equiv & \min_{x\in\{0,1\}^k} \E_{\hf_i,P_{x,\tau}}[\w_i{(f_i(x),\hf_i(\tau,x_i))}] \;.	\label{eq:wcacc}
\end{eqnarray}
For worst-case accuracy, given a protocol $P$, the optimal decision rule of the $i$-th party with a bit $x_i$
can be computed by solving the following convex program:
\begin{eqnarray}
	Q^{(x_i)} = \underset{Q \,\in\, \reals^{|\cT|\times|\cY|}}{\text{arg max}} &&
	\min_{x_{-i}\in\{0,1\}^{k-1}} \sum_{\tau\in\cT}\sum_{y\in\cY} P_{x,\tau} \,\w_i(f_i (x),y) Q_{\tau,y}
	\label{eq:optwc} \\
	\text{subject to}&&  \sum_{y\in\cY} Q_{\tau,y}=1 \;,\;\forall \tau\in\cT \text{ and } Q\geq 0 \nonumber
\end{eqnarray}
The optimal (random) decision rule $f^*_{i,\rm wc}(\tau,x_i)$ is to output $y$ given transcript $\tau$ according to $\prob(y|\tau,x_i)=Q^{(x_i)}_{\tau,y}$.
This can be formulated as a linear program with $|\cT|\,\times\,|\cY|$ variables and $2^k+|\cT|$ constraints.
Again, it is possible to find a closed form solution for the decision rule when
$f$ has a simple structure: for the XOR function,
the optimal decision rule is again evaluating the XOR of all the received bits requiring $O(k)$ operations.
For a central observer, the accuracy measures are defined similarly, and the optimal decision rule is now
\begin{eqnarray}
	f^*_{0,\rm ave}(\tau) &=& \arg\max_{y\in\cY} \sum_{x\in\{0,1\}^{k}} P_{x,\tau} \,\w_0(f_0(x),y)\;,   \label{eq:scenariotwo1}
\end{eqnarray}
and for worst-case accuracy the optimal (random) decision rule $f^*_{0,\rm wc}(\tau)$ is
to output $y$ given transcript $\tau$ according to $\prob(y|\tau)=Q^{(0)}_{\tau,y}$.
\begin{eqnarray}
	Q^{(0)} = \underset{Q \,\in\, \reals^{|\cT|\times|\cY|}}{\text{arg max}} &&
	\min_{x\in\{0,1\}^{k}} \sum_{\tau\in\cT}\sum_{y\in\cY} P_{x,\tau} \,\w_0(f_0 (x),y) Q_{\tau,y}
	\label{eq:scenariotwo2} \\
	\text{subject to}&&  \sum_{y\in\cY} Q_{\tau,y}=1 \;,\;\forall \tau\in\cT \text{ and } Q\geq 0 \nonumber
\end{eqnarray}
where $\w_0:\cY\times\cY\to\reals$ is the measure of accuracy for the central observer.

Privacy is measured by approximate differential privacy \cite{Dwo06,DMNS06}.
Since we allow heterogeneous privacy constraints,
we use $(\varepsilon_i,\delta_i)$ to denote the desired privacy level of the $i$-th party.
We say that a protocol $P$ is $(\varepsilon_i,\delta_i)$-differentially private for the $i$-th party if
for $i\in[k]$, and all  $x_i,x_i'\in\{0,1\}$, $x_{-i}\in\{0,1\}^{k-1}$, and $S\subseteq \cT$,
\begin{eqnarray}
	\prob(\tau \in S|x_i,x_{-i}) &\leq& e^{\varepsilon_i} \, \prob(\tau \in S|x_i',x_{-i}) + \delta_i \;. \label{eq:defdp}
\end{eqnarray}
This condition ensures that no adversary can infer the private data $x_i$ with high enough confidence,
no matter what auxiliary information or computational power she might.

Consider the following simple protocol known as the {\em randomized response},
which is a term first coined by \cite{War65} and commonly used in many private communications including the multi-party setting \cite{MMPT}. We will show in Section \ref{sec:main} that this is the optimal protocol that simultaneously maximizes the accuracy for all the parties.
Each party broadcasts a randomized version of its bit
denoted by $\tx_i$ such that
 \begin{eqnarray}
 	\tx_i = \left\{
	\begin{array}{rl}
		0 & \text{if } x_i=0 \text{ with probability }\delta_i \;,\\
		1 & \text{if } x_i=0 \text{ with probability }  \frac{(1-\delta_i)e^{\varepsilon_i}}{1+e^{\varepsilon_i}}\;, \\
		2 & \text{if } x_i=0 \text{ with probability }  \frac{(1-\delta_i)}{1+e^{\varepsilon_i}}\;, \\
		3 & \text{if } x_i=0 \text{ with probability } 0 \;,
	\end{array}
	\right.
 	\tx_i = \left\{
	\begin{array}{rl}
		0 & \text{if } x_i=1 \text{ with probability } 0 \;,\\
		1 & \text{if } x_i=1 \text{ with probability }  \frac{(1-\delta_i)}{1+e^{\varepsilon_i}}\;, \\
		2 & \text{if } x_i=1 \text{ with probability }  \frac{(1-\delta_i)e^{\varepsilon_i}}{1+e^{\varepsilon_i}}\;, \\
		3 & \text{if } x_i=1 \text{ with probability } \delta_i \;.
	\end{array}
	\right.
	\label{eq:rr}
 \end{eqnarray}
The reason this randomized response is optimal is that under the hypothesis testing interpretation of differential privacy,
this mechanisms achieves the largest hypothesis testing region, i.e. $\cR(\tx_i,x_i=0,x_i=1)=\cR(\varepsilon_i,\delta_i)$ as shown in Figure \ref{fig:region1}.

\subsection{Main Result}
\label{sec:main}

We show, perhaps surprisingly, that the simple randomized response presented in \eqref{eq:rr} is
the unique optimal protocol in a very general sense.
For any desired privacy level $(\varepsilon_i,\delta_i)$, and arbitrary function $f_i$, for any accuracy measure $\w_i$, and any notion of accuracy (either average or worst case), we show that the randomized response is universally optimal.

\begin{theorem}
	Let  the optimal decision rule be defined as in \eqref{eq:optave} for the average accuracy and
	\eqref{eq:optwc} for the worst-case accuracy.
	Then, for any $(\varepsilon_i,\delta_i)$, any function $f_i:\{0,1\}^k\to \cY$, and any accuracy measure $\w_i:\cY\times\cY\to\reals$
	for $i\in[k]$,
	the randomized response for given $(\varepsilon_i,\delta_i)$
	with the optimal decision function
	achieves the maximum accuracy for the $i$-th party
	among all $\{(\varepsilon_i,\delta_i)\}$-differentially private interactive protocols  and all decision rules.
	For the central observer, the randomized response with the optimal decision rule defined in \eqref{eq:scenariotwo1} and
	\eqref{eq:scenariotwo2}
	achieves the maximum accuracy among
	all $\{(\varepsilon_i,\delta_i)\}$-differentially private interactive protocols  and all decision rules for any arbitrary function $f_0$ and any measure of accuracy $\w_0$.
	\label{thm:average}
\end{theorem}

This is a strong optimality result. Every party and the central observer
can {\em simultaneously} achieve the optimal accuracy, using a universal
randomized response. Each party only needs to know its own desired level of
privacy, its own function to be computed, and its measure of accuracy.
Optimal data release and optimal decision making are naturally separated.
It is not immediate at all that such a simple non-interactive randomized response mechanism would
achieve the maximum accuracy.  The proof critically harnesses the data processing inequalities and is provided in Appendix \ref{proof_multi_party}.

\section{Proof of Theorem \ref{thm:composition}}
\label{sec:compositionproof}

We first propose a simple mechanism
and prove that the proposed  mechanism dominates over all
$(\varepsilon,\delta)$-differentially private mechanisms.
Analyzing the privacy region achieved by the
$k$-fold composition of the proposed mechanism,
we get a bound on the  privacy region under the adaptive composition.
This gives an exact characterization of privacy under composition,
since we show both converse and achievability.
We prove that no other family of  mechanisms can achieve `more degraded' privacy (converse),
and that there is a mechanism that we propose which achieves the privacy region  (achievability).

\subsection{Achievability}

We propose the following simple mechanism $\tM_i$ at the $i$-th step in the composition.
Null hypothesis ($b=0$) outcomes $X^{i,0}=M_i(D^{i,0},q_i)$'s which are independent
and identically distributed as a discrete random variable $\tX_0\sim \tP_0(\cdot)$, where
\begin{eqnarray}
	\prob(\tX_0=x) \;=\; \tP_0(x) \;\equiv\; \left\{ \begin{array}{rl} \delta &\text{ for } x=0 \;,\\
		 \frac{(1-\delta)\,e^\varepsilon}{1+e^\varepsilon} & \text{ for } x=1 \;,\\
		\frac{1-\delta}{1+e^\varepsilon} & \text{ for } x=2 \;,\\
		0 & \text{ for } x=3 \;.
		\end{array}\right.
		\label{eq:dist0}
\end{eqnarray}
Alternative hypothesis ($b=1$) outcomes
$X^{i,1}=M_i(D^{i,1},q_i)$'s are independent
and identically distributed as a discrete random variable $\tX_1\sim \tP_1(\cdot)$, where
\begin{eqnarray}
	\prob(\tX_1=x) \;=\; \tP_1 (x) \;\equiv\;  \left\{ \begin{array}{rl} 0 &\text{ for } x=0 \;,\\
		 \frac{1-\delta}{1+e^\varepsilon} & \text{ for } x=1 \;,\\
		\frac{(1-\delta)\,e^\varepsilon}{1+e^\varepsilon} & \text{ for } x=2 \;,\\
		\delta & \text{ for } x=3 \;.
		\end{array}\right.
		\label{eq:dist1}
\end{eqnarray}
In particular, the output of this mechanism does not depend on the database $D^{i,b}$
or the query $q_i$, and only depends on the hypothesis $b$.
The privacy region of a single access to this mechanism is $\cR(\varepsilon,\delta)$
in Figure \ref{fig:region1}. Hence, by Theorem \ref{thm:converse},
all $(\varepsilon,\delta)$-differentially private mechanisms are
dominated by this mechanism.

In general,
the privacy region $\cR(M,D_0,D_1)$ of any mechanism can be represented as an intersection of multiple
$\{(\tepsilon_j,\tdelta_j)\}$ privacy regions. 
For a mechanism $M$, we can compute the
$(\tepsilon_j,\tdelta_j)$ pairs representing the privacy region as follows.
Given
a null hypothesis database $D_0$,
an alternative hypothesis database $D_1$,
and a mechanism $M$ whose output space is $\cX$,
 let $P_0$ and $P_1$ denote the probability density function of the outputs
 $M(D_0)$ and
 $M(D_1)$ respectively.
 To simplify notations we assume that $P_0$ and $P_1$ are
 symmetric, i.e. there exists a permutation $\pi$ over $\cX$ such that
 $P_0(x)=P_1(\pi(x))$ and  $P_1(x)=P_0(\pi(x))$.
This ensures that we get a symmetric privacy region.

The privacy region $\cR(M,D_0,D_1)$
can be described by its  boundaries.
Since it is a convex set,
a tangent line on the  boundary with slope $-e^{\tepsilon_j}$
can be represented by
the smallest $\tdelta_j$ such that
\begin{eqnarray}
	\PFA &\geq & -e^{\tepsilon_j}\PMD + 1- {\tdelta_j}\;, \label{eq:boundary}
\end{eqnarray}
for all rejection sets (cf. Figure \ref{fig:tangent}).
Letting $S$ denote the complement of a rejection set, such that
$\PFA=1-P_0(S)$ and $\PMD=P_1(S)$,
the minimum shift $\tdelta_j$
that still ensures that the privacy region is above the line \eqref{eq:boundary}
is defined as  $\tdelta_j = d_{\tepsilon_j}(P_0,P_1)$ where
\begin{eqnarray*}
	d_\tepsilon (P_0,P_1) &\equiv& \max_{S\subseteq \cX} \Big\{P_0(S) - e^\tepsilon\,P_1(S)\Big\} \;.
\end{eqnarray*}
The privacy region of a mechanism is completely
described by the set of slopes and shifts,
$\{ (\tepsilon_j,\tdelta_j) \,:\, \tepsilon_j\in E \text{ and } \tdelta_j=d_{\tepsilon_j}(P_0,P_1) \}$,
where
\begin{eqnarray*}
	E &\equiv& \{\, 0\leq \tepsilon < \infty \,:\, P_0(x) = e^\tepsilon\,P_1(x) \text{ for some }x\in\cX \} \;.
\end{eqnarray*}
Any $\tepsilon \notin E$ does not contribute to the boundary of the privacy region.
For the above example distributions $\tP_0$ and $\tP_1$,
$E = \{\varepsilon\}$
and $d_\varepsilon(\tP_0,\tP_1)=\delta$.

\begin{remark}
	\label{lem:prob2region}
	For a database access mechanism $M$ over a output space $\cX$ and
	a pair of neighboring databases $D_0$ and $D_1$,
	let $P_0$ and $P_1$ denote the probability density function for random variables
	$M(D_0)$ and $M(D_1)$ respectively.
	Assume there exists a permutation $\pi$ over $\cX$ such that
	$P_0(x) = P_1(\pi(x))$.
	Then, the privacy region is
	\begin{eqnarray*}
		\cR(\,M,D_0,D_1\,) &=& \bigcap_{\tepsilon\in E} \cR\big(\,\tepsilon, d_\tepsilon(P_0,P_1 ) \,\big)\;,
	\end{eqnarray*}
	where $\cR(M,D,D')$ and $\cR(\tepsilon,\tdelta)$ are defined as in \eqref{eq:defregion2}
	and \eqref{eq:defregion1}.
\end{remark}
The symmetry  assumption is to simplify notations, and
the analysis can be easily generalized to deal with non-symmetric distributions.

Now consider a $k$-fold composition experiment, where
at each sequential access $\tM_i$,
we receive a random output $X^{i,b}$ independent and identically distributed as $\tX_b$.
We can explicitly characterize the distribution of $k$-fold composition of the outcomes:
$\prob(X^{1,b}=x_1,\ldots,X^{k,b}=x_k) = \prod_{x=1}^k \tP_b(x_i)$.
It follows form the structure of these two discrete distributions that,
$E=\{e^{(k-2\lfloor k /2\rfloor) \varepsilon},e^{(k+2-2\lfloor k/2\rfloor)\varepsilon},\ldots,e^{(k-2)\varepsilon},e^{k\varepsilon}\}$.
After some algebra, it also follows that
\begin{eqnarray*}
	d_{(k-2i)\varepsilon}\big(\,(\tP_0)^k,(\tP_1)^k\,\big)  &=& 1-(1-\delta)^k +(1-\delta)^k \frac{\sum_{\ell=0}^{i-1} {k \choose \ell}\big(e^{\varepsilon(k-\ell)}-e^{\varepsilon(k-2i+\ell)}\big) }{(1+e^\varepsilon)^k}\;.
\end{eqnarray*}
 for $i\in\{0,\ldots,\lfloor k/2\rfloor\}$.
 From Remark \ref{lem:prob2region}, it follows that
the privacy region is
$\cR(\{\varepsilon_i,\delta_i\}) = \bigcap_{i=0}^{\lfloor k/2\rfloor}\cR\big(\varepsilon_i,\delta_i\big)$,
where $\varepsilon_i=(k-2i)\varepsilon$ and $\delta_i$'s are defined as in \eqref{eq:defed}.
Figure \ref{fig:compose} shows this privacy region for $k=1,\ldots,5$ and for $\varepsilon=0.4$ and
for two values of $\delta=0$ and $\delta=0.1$.

\subsection{Converse}

We will now prove that this region is the largest region achievable under $k$-fold adaptive composition of
any ($\varepsilon,\delta$)-differentially private mechanisms.

From Corollary \ref{coro:hypo},
any mechanism whose privacy region is included in $\cR(\{\varepsilon_i,\delta_i\})$
satisfies $(\tepsilon,\tdelta)$-differential privacy.
We are left to prove that
for the family of all $(\varepsilon,\delta)$-differentially private mechanisms,
the privacy region of
the $k$-fold composition experiment
is included inside $\cR(\{\varepsilon_i,\delta_i\})$.
To this end, consider the following
composition experiment,
which reproduces the {\em view of the adversary} from the original composition experiment.

At each time step $i$,
we generate a random variable $X^{i,b}$ distributed as $\tX_b$
independent of any other random events,
and call this the output of a database access mechanism $\tM_i$
such that $\tM_i(D^{i,b},q_i)=X^{i,b}$.
Since, $X^{i,b}$ only depends on $b$, and
is independent of the actual database or the query,
we use $\tM_i(b)$ to denote this outcome.

We know that
$\tM_i(b)$ has privacy region $\cR(\varepsilon,\delta)$
for any choices of $D^{i,0}$, $D^{i,1}$ and $q_i$.
Now consider the mechanism $M_i$ from the original experiment.
Since it is $(\varepsilon,\delta)$-differentially private,
we know from Theorem \ref{thm:hypo} that
$\cR(M_i,D^{i,0},D^{i,1}) \subseteq\cR(\varepsilon,\delta)$ for
any choice of neighboring databases  $D^{i,0}$, $D^{i,1}$.
Hence, from the converse of data processing inequality (Theorem \ref{thm:converse}),
we know that there exists a mechanism $T_i$ that takes
as input $X^{i,b}$ 
and produces an output $Y^{i,b}$ which is distributed as
$M_i(D^{i,b},q_i)$ for all $b\in\{0,1\}$. Hence,
$Y^{i,b}$ is independent of the past
conditioned on $X^{i,b},D^{i,0},D^{i,1},q_i,M_i$.
Precisely we have the following Markov chain:
\begin{eqnarray*}
	(b,R,\{X^{\ell,b},D^{\ell,0},D^{\ell,1},q_\ell,M_\ell\}_{\ell\in[i-1]})\text{--}(X^{i,b},D^{i,0},D^{i,1},q_i,M_i) \text{--}Y^{i,b}\;,
\end{eqnarray*}
where $R$ is any internal randomness of the adversary $\cA$.
Since, $(X,Y)$--$Z$--$W$ implies $X$--$(Y,Z)$--$W$, we have
\begin{eqnarray*}
	b\text{--}(R,\{X^{\ell,b},D^{\ell,0},D^{\ell,1},q_\ell,M_\ell\}_{\ell\in[i]}) \text{--}Y^{i,b}\;.
\end{eqnarray*}
Notice that if we know $R$ and the outcomes $\{Y^{\ell,b}\}_{\ell\in[i]}$,
then we can reproduce the original experiment until time $i$.
This is because the choices of  $D^{i,0},D^{i,1},q_i,M_i$ are exactly specified
by  $R$ and $\{Y^{\ell,b}\}_{\ell\in[i]}$. Hence, we can simplify the Markov chain as
\begin{eqnarray}
	b\text{--}(R,X^{i,b},\{X^{\ell,b},Y^{\ell,b}\}_{\ell\in[i-1]}) \text{--}Y^{i,b}\;.
	\label{eq:markov1}
\end{eqnarray}
Further, since $X^{i,b}$ is independent of the past conditioned on $b$,
we have
\begin{eqnarray}
	X^{i,b}\text{--}b\text{--}(R,\{X^{\ell,b},Y^{\ell,b}\}_{\ell\in[i-1]}) \;.   \label{eq:markov2}
\end{eqnarray}

It follows that
\begin{align*}
&\prob(b,r,x_1\ldots,x_k,y_1,\ldots,y_k)
\;\;= \;\;\prob(b,r,x_1,\ldots,x_k,y_1,\ldots,y_{k-1}) \prob(y_k|r,x_1,\ldots,x_k,y_1,\ldots,y_{k-1})\\
&\;\;\;\;\;=\;\; \prob(b,r,x_1,\ldots,x_{k-1},y_1,\ldots,y_{k-1}) \prob(x_k|b) \prob(y_k|r,x_1,\ldots,x_k,y_1,\ldots,y_{k-1})\;,
\end{align*}
where we used \eqref{eq:markov1} in the first equality and \eqref{eq:markov2} in the second.
By induction, we get a decomposition
\begin{eqnarray*}
	\prob(b,r,x_1,\ldots,x_k,y_1,\ldots,y_k) & =& \prob(b,r) \prod_{i=1}^k \prob(x_i|b) \prod_{i=1}^k \prob(y_i|r,x_1,\ldots,x_i,y_1,\ldots,y_{i-1}) \\
	&=& \prob(b,r,x_1,\ldots,x_k) \prob(y_1,\ldots,y_k| r,  x_1,\ldots,x_k ) \\
	&=& \prob(b|r,x_1,\ldots,x_k)\,\prob(y_1,\ldots,y_k,r,x_1,\ldots,x_k) \;.
\end{eqnarray*}
From the construction of the experiment, it also follows that
the internal randomness $R$ is independent of the
hypothesis $b$ and the outcomes $X^{i,b}$'s:
$\prob(b|r,x_1,\ldots,x_k)=\prob(b|x_1,\ldots,x_k)$.
Then, marginalizing over $R$, we get
$\prob(b,x_1,\ldots,x_k,y_1,\ldots,y_k) =\prob(b|x_1,\ldots,x_k)\,\prob(y_1,\ldots,y_k,x_1,\ldots,x_k)$.
This implies the following Markov chain:
\begin{eqnarray}
	b\text{--}(\{X^{i,b}\}_{i\in[k]})\text{--}(\{Y^{i,b}\}_{i\in[k]}) \;,
	\label{eq:markov3}
\end{eqnarray}
and it follows that a set of mechanisms $(M_1,\ldots,M_k)$ dominates
$(\tM_1,\ldots,\tM_k)$ for two databases $\{D^{i,0}\}_{i\in[k]}$ and $\{D^{i,1}\}_{i\in[k]}$.
By the data processing inequality for differential privacy (Theorem \ref{thm:dpi}),
this implies that
\begin{eqnarray*}
	\cR\big(\{M_i\}_{i\in[k]},\{D^{i,0}\}_{i\in[k]},\{D^{i,1}\}_{i\in[k]}\big) &\subseteq&
		\cR\big(\{\tM_i\}_{i\in[k]},\{D^{i,0}\}_{i\in[k]},\{D^{i,1}\}_{i\in[k]}\big) \;=\; \cR\big(\{\varepsilon_i,\delta_i\}\big)\;.
\end{eqnarray*}
This finishes the proof of the desired claim.

Alternatively, one can prove \eqref{eq:markov3}, using a probabilistic graphical model.
Precisely, the following Bayesian network describes the dependencies among various
random quantities of the experiment described above.
Since the set of nodes $(X^{1,b},X^{2,b},X^{3,b},X^{4,b})$ d-separates node $b$ from the rest of the bayesian network,
it follows immediately from the Markov property of this Bayesian network that
\eqref{eq:markov3} is true (cf. \cite{Lau96}).
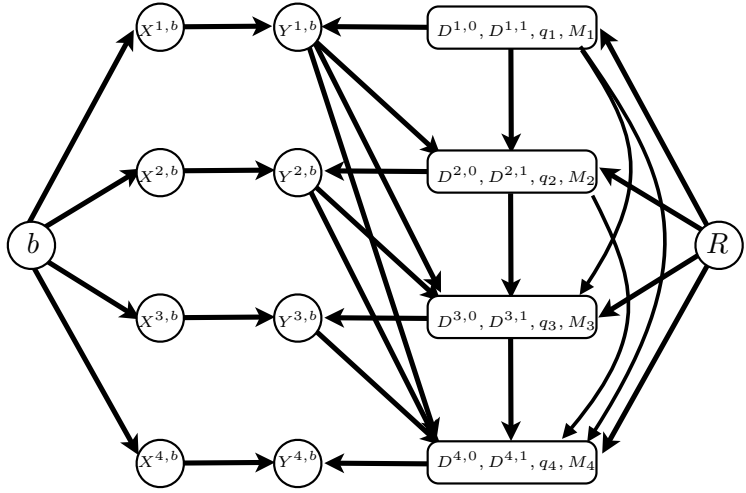
\begin{figure}[h]
\begin{center}
	\includegraphics[width=.6\textwidth]{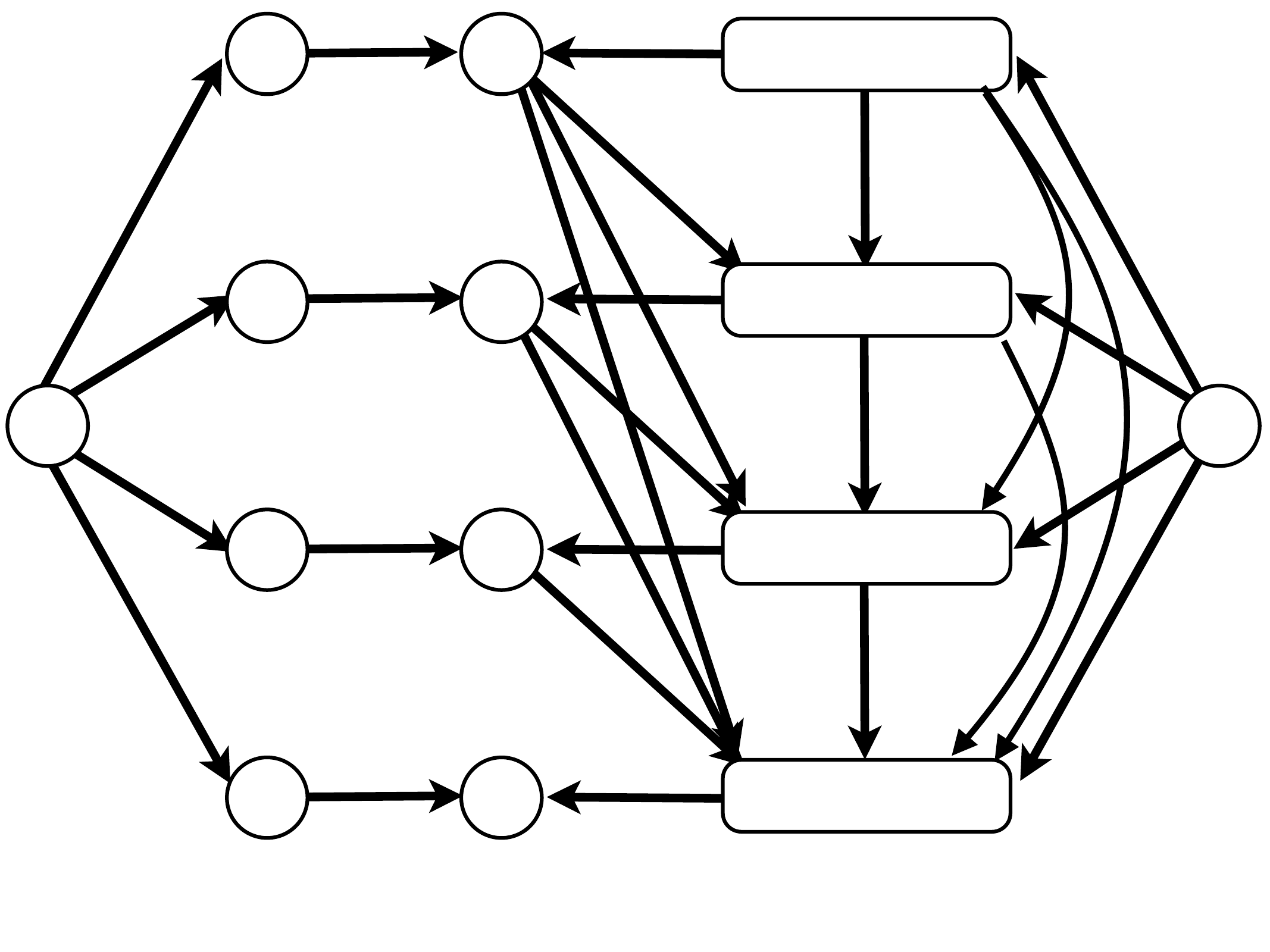}
	\put(-273,114){$b$}
	\put(-16,114){$R$}
	\put(-231,196){\tiny$X^{1,b}$}
	\put(-231,141){\tiny$X^{2,b}$}
	\put(-231,86){\tiny$X^{3,b}$}
	\put(-231,33){\tiny$X^{4,b}$}
	\put(-178,196){\tiny$Y^{1,b}$}
	\put(-178,141){\tiny$Y^{2,b}$}
	\put(-178,86){\tiny$Y^{3,b}$}
	\put(-178,33){\tiny$Y^{4,b}$}
	\put(-118,196){\tiny$D^{1,0},D^{1,1},q_1,M_1$}
	\put(-118,141){\tiny$D^{2,0},D^{2,1},q_2,M_2$}
	\put(-118,86){\tiny$D^{3,0},D^{3,1},q_3,M_3$}
	\put(-118,33){\tiny$D^{4,0},D^{4,1},q_4,M_4$}
\end{center}
	\caption{Bayesian network representation of the composition experiment.
	The subset of nodes $(X^{1,b},X^{2,b},X^{3,b},X^{4,b})$ d-separates node $b$ from the rest of the network.}
\end{figure}

%
%
\section{Proof of Theorem \ref{thm:closedform}}
\label{sec:closedform}

We need to provide an outer bound on the privacy region achieved by
$\tX_0$ and $\tX_1$ defined in \eqref{eq:dist0} and \eqref{eq:dist1}
under $k$-fold composition.
Let $P_0$ denote the probability mass function of $\tX_0$
and $P_1$ denote the PMF of $\tX_1$.
Also, let $P_0^k$ and $P_1^k$
denote the joint PMF of $k$ i.i.d. copies of $\tX_0$ and $\tX_1$ respectively.
Also, for a set $S\subseteq \cX^k$, we let $P_0^k(S)=\sum_{x\in S} P_0^k(x)$.
In our example, $\cX=\{1,2,3,4\}$, and
\begin{eqnarray*}
	P_0 &=& \begin{bmatrix} \delta & \frac{(1-\delta)e^\varepsilon}{1+e^\varepsilon} & \frac{1-\delta}{1+e^\varepsilon} & 0 \end{bmatrix}\;, \\
	P_1 &=& \begin{bmatrix} 0& \frac{1-\delta}{1+e^\varepsilon} & \frac{(1-\delta)e^\varepsilon}{1+e^\varepsilon} & \delta\end{bmatrix}\;, \\
	P_0^2 &=& \begin{bmatrix}
		\delta^2 &  \delta\frac{(1-\delta)e^\varepsilon}{1+e^\varepsilon} & \delta \frac{(1-\delta)}{1+e^\varepsilon}  & 0\\
		 \delta \frac{(1-\delta)e^\varepsilon}{1+e^\varepsilon} & \Big(\frac{(1-\delta)e^\varepsilon}{1+e^\varepsilon}\Big)^2 & \Big(\frac{1-\delta}{1+e^\varepsilon}\Big)^2e^\varepsilon & 0\\
		 \delta \frac{1-\delta}{1+e^\varepsilon} &\Big(\frac{1-\delta}{1+e^\varepsilon}\Big)^2e^\varepsilon & \Big(\frac{1-\delta}{1+e^\varepsilon}\Big)^2 & 0\\
		0 &0 &0 & 0
	\end{bmatrix}\;, \text{etc.}
\end{eqnarray*}

We can compute the privacy region from $P_0^k$ and $P_1^k$ directly,
by computing the line tangent to the boundary.
A tangent line with slope $-e^\tepsilon$ can be represented as
\begin{eqnarray}
	\PFA = -e^\tepsilon \PMD + 1- d_\tepsilon(P_0^k,P_1^k)\;. \label{eq:tangent}
\end{eqnarray}
To find the tangent line, we need to maximize the shift, which is equivalent to moving the line downward until it is tangent to the boundary of the privacy region
(cf. Figure \ref{fig:tangent}).
\begin{eqnarray*}
	d_\tepsilon(P_0^k,P_1^k) &\equiv&
		\max_{S\subseteq \cX^k} \, P_0^k(S) - e^\tepsilon  P_1^k(S) \;.
\end{eqnarray*}
Notice that the maximum is achieved by a set
$B \equiv \{ x \in \cX^k \,|\, P_0^k(x)\geq e^{\tepsilon} P_1^k(x)\}$.  Then,
\begin{eqnarray*}
	d_\tepsilon(P_0^k,P_1^k)  &=& P_0^k(B)-e^\tepsilon P_1^k(B) \;.
\end{eqnarray*}

For the purpose of proving the bound of the form \eqref{eq:closedform},
we separate the analysis of the above formula into two parts:
one where either $P_0^k(x)$ or $P_1^k(x)$ is zero and
the other when both are positive.
Effectively, this separation allows us to
treat the effects of $(\varepsilon,0)$-differential privacy and
$(0,\delta)$-differential privacy separately.
In previous work \cite{DRV10}, 
they separated the analysis in a similar way.
Here we provide a simpler proof technique.
Further, all the proof techniques we use
naturally generalize to compositions of general $(\varepsilon,\delta)$-differentially private
mechanisms other than
the specific example of $\tX_0$ and $\tX_1$ we consider in this section.

Let $\tX^k_0$ denote a $k$-dimensional random vector whose entries are
independent copies of $\tX_0$.
We partition $B$ into two sets: $B=B_0\bigcup B_1$ and $B_0 \bigcap B_1 = \emptyset$.
Let $B_0 \equiv  \{ x \in \cX^k \,:\, P_0^k(x)\geq e^{\tepsilon} P_1^k(x) \text{, and } P_1^k(x)=0\} $
and $B_1 \equiv  \{ x \in \cX^k \,:\, P_0^k(x)\geq e^{\tepsilon} P_1^k(x) \text{, and } P_1^k(x)>0 \} $.
Then, it is not hard to see that
$P_0^k(B_0) = 1- \prob(\tX_0^k \in \{1,2,3\}^k) = 1-(1-\delta)^k$,
$P_1^k(B_0) = 0$,
$P_0^k(B_1) = P_0^k(B_1|\tX_0^k \in \{1,2\}^k)  \prob(\tX_0^k \in \{1,2\}^k) =
(1-\delta)^k\,P_0^k(B_1|\tX_0^k \in \{1,2\}^k)$, and
$P_1^k(B_1) = (1-\delta)^k \, P_1^k(B_1|\tX_1^k \in \{1,2\}^k)$. It follows that
\begin{eqnarray*}
	P_0^k(B_0)-e^\tepsilon P_1^k(B_0) &=& 1- (1-\delta)^k\;, \text{ and }\\
	P_0^k(B_1)-e^\tepsilon P_1^k(B_1) &=& (1-\delta)^k\big( P_0^k(B_1|\tX_0^k \in \{1,2\}^k)-e^\tepsilon P_1^k(B_1|\tX_1^k \in \{1,2\}^k)\big)\;.
\end{eqnarray*}
 Let $\tP_0^k (x)  \equiv P_0^k(x|x\in\{1,2\}^k)$ and
 $\tP_1^k (x)  \equiv P_1^k (x|x\in\{1,2\}^k)$. Then, we have
\begin{eqnarray}
	d_\tepsilon(P_0^k,P_1^k) &=&	P_0^k(B_0)-e^\tepsilon P_1^k(B_0) + P_0^k(B_1)-e^\tepsilon P_1^k(B_1) \nonumber\\
		&=& 1- (1-\delta)^k + (1-\delta)^k \big( \tP_0^k(B_1)-e^\tepsilon \tP_1^k(B_1)\big)\;. \label{eq:deltabound}
\end{eqnarray}
Now, we focus on upper bounding
$\tP_0^k(B_1)-e^\tepsilon \tP_1^k(B_1)$, using a variant of Chernoff's tail bound.
Notice that
\begin{eqnarray}
	\tP_0^k(B_1)-e^\tepsilon \tP_1^k(B_1) &=& \E_{\tP_0^k} \big[ \ind_{\big(\log(\tP_0^k(\tX^k)/\tP_1^k(\tX^k))\geq \tepsilon\big)}  \big] -e^{\tepsilon}  \E_{\tP_0^k} \big[ \ind_{\big(\log(\tP_0^k(\tX^k)/\tP_1^k(\tX^k))\geq \tepsilon\big)} \frac{\tP_1^k(\tX^k)}{\tP_0^k(\tX^k) } \big] \nonumber\\
		&=& \E_{\tP_0^k} \Big[\ind_{\big(\log(\tP_0^k(\tX^k)/\tP_1^k(\tX^k))\geq \tepsilon\big)} \Big(1-e^\tepsilon \frac{\tP_1^k(\tX^k)}{\tP_0^k(\tX^k)} \Big) \Big] \nonumber \\
		&\leq& \E [e^{\lambda Z- \lambda\tepsilon+\lambda\log\lambda-(\lambda+1)\log(\lambda+1)}] \;, \label{eq:deltabound2}
\end{eqnarray}
where we use a random variable $Z \equiv \log(\tP_0^k(\tX_0^k) / \tP_1^k(\tX_0^k))$
and the last line follows from
$\ind_{(x\geq\tepsilon)}(1-e^{\tepsilon -x}) \leq e^{\lambda(x-\tepsilon)+\lambda\log\lambda-(\lambda+1)\log(\lambda+1)}$ for any $\lambda\geq0$.
To show this inequality, notice that
the right-hand side is always non-negative.
So it is sufficient to show that the inequality holds, without the indicator on the left-hand side.
Precisely, let
$f(x)=e^{\lambda(x-\tepsilon)+\lambda\log\lambda-(\lambda+1)\log(\lambda+1)} + e^{\tepsilon-x} -1$.
This is a convex function with $f(x^*)=0$ and $f'(x^*)=0$ at
$x^*=\tepsilon+\log((\lambda+1)/\lambda)$. It follows that this is a non-negative function.

Next, we give an upper bound on the moment generating function of $Z$.
\begin{eqnarray*}
	\E_{\tP_0}[e^{\lambda \log(P_0(X)/P_1(X))}] &=& \frac{e^\varepsilon}{e^\varepsilon+1} e^{\lambda\varepsilon} + \frac{1}{e^\varepsilon+1}e^{-\lambda\varepsilon} \\
	&\leq& e^{\frac{e^\varepsilon-1}{e^\varepsilon+1}\lambda\varepsilon + \frac12\lambda^2\varepsilon^2}\;,
\end{eqnarray*}
for any $\lambda$, which follows from the fact that
$ p e^x + (1-p)e^{-x} \leq e^{(2p-1)x+(1/2)x^2}$ for any $x\in\reals$ and $p\in[0,1]$ \cite[Lemma A.1.5]{AS04}.
Substituting this into \eqref{eq:deltabound2} with a
choice of $\lambda=\frac{\tepsilon - k\varepsilon(e^\varepsilon-1)/(e^\varepsilon+1)}{k\varepsilon^2}$, we get
\begin{align*}
&	\tP_0^k(B_1)-e^\tepsilon \tP_1^k(B_1)  \leq
	\exp\Big\{ \frac{e^\varepsilon-1}{e^\varepsilon+1}  \lambda\varepsilon k
		+ \frac12 \lambda^2 \varepsilon^2 k -\lambda\tepsilon +   \lambda\log\lambda-(\lambda+1)\log(\lambda+1) \Big\}\\
		&\hspace{1cm}= \exp\Big\{ -\frac{k \varepsilon^2}{2} \Big(\lambda- \frac{1}{k\varepsilon^2} \Big( \tepsilon-k\varepsilon\frac{e^\varepsilon-1}{e^\varepsilon+1}\Big)\, \Big)^2 \, -\frac{1}{2k\varepsilon^2}\Big(\tepsilon-\frac{k\varepsilon(e^\varepsilon-1)}{e^\varepsilon+1} \Big)^2+ \lambda\log\frac{\lambda}{\lambda+1}- \, \log(\lambda+1) \Big\} \\
		&\hspace{1cm}\leq \exp\Big\{ -\frac{1}{2 k \varepsilon^2}\Big(\tepsilon-k\varepsilon\frac{e^\varepsilon-1}{e^\varepsilon+1}\Big)^2 \, - \, \log(\lambda+1) \Big\} \\
		&\hspace{1cm}\leq \frac{1}{1+\frac{\tepsilon-k\varepsilon(e^\varepsilon-1)/(e^\varepsilon+1)}{k\varepsilon^2}} \exp\Big\{ -\frac{1}{2 k \varepsilon^2}\Big(\tepsilon-k\varepsilon\frac{e^\varepsilon-1}{e^\varepsilon+1}\Big)^2\, \Big\} \\
		&\hspace{1cm} = \frac{1}{1+\frac{\sqrt{2k\varepsilon^2 \log(e+(\sqrt{k\varepsilon^2}/\tdelta))}}{k\varepsilon^2}} \frac{1}{e+\frac{\sqrt{k \varepsilon^2}}{\tdelta}} \\
		& \hspace{1cm} \leq \frac{1}{\sqrt{k \varepsilon^2}+\sqrt{2 \log(e+(\sqrt{k\varepsilon^2}/\tdelta))}} \frac{\tdelta}{\frac{e\tdelta}{\sqrt{k \varepsilon^2}}+1} \;,
\end{align*}
for our choice of
$\tepsilon=k\varepsilon(e^\varepsilon-1)/(e^\varepsilon+1) + \varepsilon\sqrt{2k\,\log(e+(\sqrt{k\varepsilon^2}/\tdelta))}$.
The right-hand side is always less than $\tdelta$.

Similarly, one can show that the right-hand side is less than $\tdelta$ for the choice of
$\tepsilon = k\varepsilon(e^\varepsilon-1)/(e^\varepsilon+1) + \varepsilon\sqrt{2k\,\log(1/\tdelta)}$.
We get that the $k$-fold composition is
$(\tepsilon,1-(1-\delta)^k (1-\tdelta))$-differentially private.


%
%
\section{Proof of Theorem \ref{thm:hetero}}
\label{sec:hetero}

In this section, we closely follow the proof of Theorem \ref{thm:closedform} in Section \ref{sec:closedform}
carefully keeping the dependence on $\ell$, the index of the composition step.
For brevity, we omit the details which overlap with the proof of Theorem
\ref{thm:closedform}.
By the same argument as in the proof of
Theorem \ref{thm:composition}, we only need to provide an outer bound on the privacy region achieved by
$\tX_0^{(\ell)}$ and $\tX_1^{(\ell)}$ under $k$-fold composition, defined as
\begin{eqnarray*}
	\prob(\tX_0^{(\ell)}=x) \;=\; \tP_0^{(\ell)}(x) \;\equiv\; \left\{ \begin{array}{rl} \delta_\ell &\text{ for } x=0 \;,\\
		 \frac{(1-\delta_\ell)\,e^{\varepsilon_\ell}}{1+e^{\varepsilon_\ell}} & \text{ for } x=1 \;,\\
		\frac{1-\delta_\ell}{1+e^{\varepsilon_\ell}} & \text{ for } x=2 \;,\\
		0 & \text{ for } x=3 \;.
		\end{array}\right. \;, \text{ and }
\end{eqnarray*}
\begin{eqnarray*}
	\prob(\tX_1^{(\ell)}=x) \;=\; \tP_1^{(\ell)} (x) \;\equiv\;  \left\{ \begin{array}{rl} 0 &\text{ for } x=0 \;,\\
		 \frac{1-\delta_\ell}{1+e^{\varepsilon_\ell}} & \text{ for } x=1 \;,\\
		\frac{(1-\delta_\ell)\,e^{\varepsilon_\ell}}{1+e^{\varepsilon_\ell}} & \text{ for } x=2 \;,\\
		\delta_\ell & \text{ for } x=3 \;.
		\end{array}\right.
\end{eqnarray*}
Using the similar notations as Section \ref{sec:closedform},
it follows that under $k$-fold composition,
\begin{eqnarray}
	d_\tepsilon(P_0^k,P_1^k)
		&=& 1- \prod_{\ell=1}^k (1-\delta_\ell) + \big( \tP_0^k(B_1)-e^\tepsilon \tP_1^k(B_1)\big) \prod_{\ell=1}^k (1-\delta_\ell) \;. \label{eq:heterobound}
\end{eqnarray}
Now, we focus on upper bounding
$\tP_0^k(B_1)-e^\tepsilon \tP_1^k(B_1)$, using a variant of Chernoff's tail bound.
We know that
\begin{eqnarray}
	\tP_0^k(B_1)-e^\tepsilon \tP_1^k(B_1) &=& \E_{\tP_0^k} \big[ \ind_{\big(\log(\tP_0^k(\tX^k)/\tP_1^k(\tX^k))\geq \tepsilon\big)}  \big] -e^{\tepsilon}  \E_{\tP_0^k} \big[ \ind_{\big(\log(\tP_0^k(\tX^k)/\tP_1^k(\tX^k))\geq \tepsilon\big)} \frac{\tP_1^k(\tX^k)}{\tP_0^k(\tX^k) } \big] \nonumber\\
		&=& \E_{\tP_0^k} \Big[\ind_{\big(\log(\tP_0^k(\tX^k)/\tP_1^k(\tX^k))\geq \tepsilon\big)} \Big(1-e^\tepsilon \frac{\tP_1^k(\tX^k)}{\tP_0^k(\tX^k)} \Big) \Big] \nonumber \\
		&\leq& \E [e^{\lambda Z- \lambda\tepsilon+\lambda\log\lambda-(\lambda+1)\log(\lambda+1)}] \;, \label{eq:heterobound2}
\end{eqnarray}
where we use a random variable $Z \equiv \log(\tP_0^k(\tX_0^k) / \tP_1^k(\tX_0^k))$
and the last line follows from the fact that
$\ind_{(x\geq\tepsilon)}(1-e^{\tepsilon -x}) \leq e^{\lambda(x-\tepsilon)+\lambda\log\lambda-(\lambda+1)\log(\lambda+1)}$ for any $\lambda\geq0$.

Next, we give an upper bounds on the moment generating function of $Z$.
From the definition of $\tP_0^{(\ell)}$ and $\tP_1^{(\ell)}$,
$\E[e^{\lambda Z}]= \Big(\E_{\tP_0^{(\ell)}}[e^{\lambda \log(\tP_0^{(\ell)}(\tX_0^{(\ell)})/\tP_1^{(\ell)}(\tX_0^{(\ell)}))}]\Big)^k$.
Let $\tepsilon = \sum_{\ell=1}^k (e^{\varepsilon_\ell}-1)\varepsilon_\ell/(e^{\varepsilon_\ell}+1) + \sqrt{2\sum_{\ell=1}^k \varepsilon_\ell^2 \log\big(e + (\sqrt{\sum_{\ell=1}^k \varepsilon_\ell^2}/\tdelta)\big) }$.
Next we show that
the k-fold composition is $(\tepsilon,1-(1-\tdelta)\prod_{\ell\in[k]} (1-\delta_\ell)\,)$-differentially private.
\begin{eqnarray*}
	\E_{\tP_0^{(\ell)}}[e^{\lambda \log(P_0^{(\ell)}(X)/P_1^{(\ell)}(X))}]
	&\leq& e^{\frac{e^{\varepsilon_\ell}-1}{e^{\varepsilon_\ell}+1}\lambda{\varepsilon_\ell} + \frac12\lambda^2{\varepsilon_\ell}^2}\;,
\end{eqnarray*}
for any $\lambda$.
Substituting this into \eqref{eq:heterobound2} with a
choice of $\lambda=\frac{\tepsilon - \sum_{\ell\in[k]}\varepsilon_\ell(e^{\varepsilon_\ell}-1)/(e^{\varepsilon_\ell}+1)}{\sum_{\ell\in[k]}\varepsilon_\ell^2}$, we get
\begin{eqnarray*}
	\tP_0^k(B_1)-e^\tepsilon \tP_1^k(B_1)
		&\leq& \frac{1}{1+\frac{\tepsilon - \sum_{\ell\in[k]}\varepsilon_\ell(e^{\varepsilon_\ell}-1)/(e^{\varepsilon_\ell}+1)}{\sum_{\ell\in[k]}\varepsilon_\ell^2}}\exp\Big\{ -\frac{1}{2 \sum_{\ell\in[k]} \varepsilon_\ell^2}\Big(\tepsilon-\sum_{\ell\in[k]}\varepsilon_\ell\frac{e^{\varepsilon_\ell}-1}{e^{\varepsilon_\ell}+1}\Big)^2 \,\Big\} \;. \\
\end{eqnarray*}
Substituting $\tepsilon$, we get the desired bound.

Similarly, we can prove that with
$\tepsilon = \sum_{\ell=1}^k (e^{\varepsilon_\ell}-1)\varepsilon_\ell/(e^{\varepsilon_\ell}+1) + \sqrt{2\sum_{\ell=1}^k \varepsilon_\ell^2 \log\big(1/\tdelta \big) }$, the desired bound also holds.
%
%
\section{Proofs}
\subsection{Proof of Theorem \ref{thm:dpi}}
\label{sec:proofdpi}
Consider hypothesis testing between $D_1$ and $D_2$.
If there is a point $(\PMD,\PFA)$ achieved by $M'$ but not by $M$,
then we claim that this is a contradiction to
the assumption that $D$--$X$--$Y$ form a Markov chain.
Consider a decision maker who have only access to the output of $M$.
Under the Markov chain assumption,
he can simulate the output of $M'$
by generating
a random variable $Y$ conditioned on $M(D)$
and achieve every point in the privacy region of $M'$ (cf. Theorem \ref{rem:convexhull}).
Hence, the privacy region of $M'$ must be included in the privacy region of $M$.

\subsection{Proof of Theorem \ref{thm:hypo}}
\label{sec:hypo}
First we prove that $(\varepsilon,\delta$)-differential privacy implies \eqref{eq:hypo}.
From the definition of differential privacy,
we know that for all rejection set $S\subseteq \cX$,
$\prob(M(D_0) \in \bS) \leq e^\varepsilon \prob(M(D_1)\in\bS) + \delta$.
This implies
$1-\PFA(D_0,D_1,M,S) \leq e^\varepsilon \PMD(D_0,D_1,M,S) + \delta$.
This implies the first inequality of \eqref{eq:hypo}, and the second one follows similarly.

The converse follows analogously.
For any set $S$, we assume
$1-\PFA(D_0,D_1,M,S) \leq e^\varepsilon \PMD(D_0,D_1,M,S) + \delta$.
Then,
it follows that
$\prob(M(D_0) \in \bS) \leq e^\varepsilon \prob(M(D_1)\in\bS) + \delta$
for all choices of $S\subseteq \cX$. Together with the symmetric condition
$\prob(M(D_1) \in \bS) \leq e^\varepsilon \prob(M(D_0)\in\bS) + \delta$ ,
this implies $(\varepsilon,\delta)$-differential privacy.

%
%
\subsection{Proof of Remark \ref{rem:convexhull}}
\label{sec:convexhull}

We have a decision rule $\gamma$ represented by a partition $\{S_i\}_{i\in\{1,\ldots,N\}}$ and corresponding
accept probabilities $\{p_i\}_{i\in\{1,\ldots,N\}}$, such that
if the output is in a set $S_i$, we accept with probability $p_i$.
We assume the subsets are sorted such that $1\geq p_1 \geq \ldots \geq p_N\geq0$.
Then, the probability of false alarm is
\begin{eqnarray*}
	\PFA(D_0,D_1,M,\gamma) &=& \sum_{i=1}^N p_i \,\prob(M(D_0)\in  S_i)
		\;\;=\;\; p_N + \sum_{i=2}^{N} (p_{i-1}-p_i)\, \prob(M(D_0)\in \cup_{j< i} S_j)\;.
\end{eqnarray*}
 and similarly,
$\PMD(D_0,D_1,M,\gamma) = (1-p_1) + \sum_{i=2}^{N} (p_{i-1}-p_{i})\, \prob(M(D_1)\notin \cup_{j < i} S_j)$.
Recall that $\PFA(D_0,D_1,M,S) = \prob( M(D_0)\in S)$ and
$\PMD(D_0,D_1,M,S) = \prob(M(D_1)\in \bS)$.
So for any decision rule $\gamma$, we can represent the pair $(\PMD,\PFA)$ as a convex combination:
\begin{eqnarray*}
	\big(\,\PMD(D_0,D_1,M,\gamma),\PFA(D_0,D_1,M,\gamma)\,\big) &=& \sum _{i=1}^{N+1} ({p_{i-1}-p_i}) \big(\, \PMD(D_0,D_1,M, \cup_{j < i} S_j),\PFA(D_0,D_1,M, \cup_{j < i} S_j)\,\big)\;,
\end{eqnarray*}
where we used $p_0 = 1$ and $p_{N+1}=0$,
and hence it is included in the convex hull of the privacy region achieved by
decision rules with hard thresholding.


\section{Acknowledgement}

The authors  thank  Maxim Raginsky for helpful discussions and for pointing out \cite{Bla53}, and Moritz Hardt for pointing out an error in an earlier version of this paper. 
This research is supported in part by NSF CISE award CCF-1422278, NSF SaTC award
CNS-1527754, NSF CMMI award MES-1450848 and NSF ENG award ECCS-1232257.

%
%
\appendix

%
%
\section{Examples illustrating the strengths of graphical representation of differential privacy}
\label{sec:simpleproof}

\begin{remark} The following statements are true.
	\begin{itemize}
		\item[(a)] If a mechanism is $(\varepsilon,\delta)$-differentially private, then it is
		$(\tepsilon,\tdelta)$-differentially private for all pairs of $\tepsilon$ and $\tdelta\geq \delta $ satisfying
		\begin{eqnarray*}
			\frac{1-\delta}{1+e^{\varepsilon}} &\geq& \frac{1-\tdelta}{1+e^{\tepsilon}}\;.
		\end{eqnarray*}
		\item[(b)] For a pair of neighboring databases $D$ and $D'$, and all $(\varepsilon,\delta)$-differentially
		 private mechanisms, the total variation distance defined as
		  $\|M(D)-M(D')\|_{\rm TV} = \max_{S\subseteq\cX} \prob(M(D')\in S)-\prob(M(D)\in S)$ is bounded by
		\begin{eqnarray*}
			\sup_{\text{$(\varepsilon,\delta)$-differentially private $M$}} \|M(D)-M(D')\|_{\rm TV} &\leq& 1 - \frac{2(1-\delta)}{1+e^{\varepsilon}}\;.
		\end{eqnarray*}
	\end{itemize}
\end{remark}
\begin{proof}
{\bf Proof of ($a$).}
From Figure~\ref{fig:region1}, it is immediate that $\cR(\varepsilon,\delta)\subseteq\cR(\tepsilon, \tdelta)$
when the conditions are satisfied. Then, for a $(\varepsilon,\delta)$-private $M$,
it follows from $\cR(M)\subseteq\cR(\varepsilon,\delta)\subseteq\cR(\tepsilon, \tdelta)$
that $M$ is $(\tepsilon,\tdelta)$-differentially private.

\bigskip\noindent
{\bf Proof of ($b$).}
By definition, $\|M(D)-M(D')\|_{\rm TV} = \max_{S\subseteq\cX} \prob(M(D')\in S)-\prob(M(D)\in S)$.
Letting $S$ be the rejection region in our hypothesis testing setting,
the total variation distance is defined by the following optimization problem:
\begin{eqnarray}
	\max_S && 1-\PMD(S)-\PFA(S) \label{eq:max}\\
	\text{subject to} && (\PMD(S),\PFA(S)) \in \cR(\varepsilon,\delta) \text{, for all $S\subseteq\cX$} \;.\nonumber
\end{eqnarray}
From Figure~\ref{fig:region1} it follows immediately that the total variation distance cannot be larger than
$\delta+(1-\delta)(e^\varepsilon-1)/(e^\varepsilon+1)$.
\end{proof}

%
%
\section{Analysis of the Gaussian mechanism in Theorem \ref{thm:Gaussian}}
\label{sec:Gaussianproof}

Following the analysis in Section \ref{sec:closedform},
we know that the privacy region of a composition of mechanisms is described by
a set of $(\varepsilon,\delta)$ pairs that satisfy the following:
\begin{eqnarray*}
	\delta   &=& \mu_0^k(B)-e^\varepsilon \mu_1^k(B) \;,
\end{eqnarray*}
where $\mu_0^k$ and $\mu_1^k$ are probability measures of the mechanism under $k$-fold composition
when the data base is $D_0$ and $D_1$ respectively,
and the subset $B = \arg\max_{S\subseteq \reals^k}  \mu_0^k(S)-e^\varepsilon \mu_1^k(S) $.

In the case of Gaussian mechanisms,
we can assume without loss of generality that
$D_0$ is such that $q_i(D_0)=0$ and
$D_1$ is such that $q_i(D_1)=\Delta$ for all $i\in\{1,\ldots,k\}$.
When adding Gaussian noises with variances $\sigma^2$,
we want to ask how small the variance can be and still ensure $(\varepsilon,\delta)$-differential privacy under $k$-fold composition.

Let $f_0^k(x_1,\ldots,x_k) =\prod_{i=1}^k f_0(x_i)=(1/\sqrt{2\pi\sigma^2})^k e^{-\sum_{i=1}^k x_i^2/2\sigma^2}$ and
$f_1^k(x_1\ldots,x_k)=\prod_{i=1}^k f_1(x_i)=(1/\sqrt{2\pi\sigma^2})^k e^{-\sum_{i=1}^k (x_i-\Delta)^2/2\sigma^2}$
be the probability density functions of Gaussians centered at zero and $\Delta\ones_k$ respectively.
Using a similar technique as in \eqref{eq:deltabound2},
we know that
\begin{eqnarray}
	\mu_0^k(B)-e^\varepsilon \mu_1^k(B) &=& \E_{\mu_0^k} \big[ \ind_{\big(\log(f_0^k(\tX^k)/f_1^k(\tX^k))\geq \varepsilon\big)}  \big] -e^{\varepsilon}  \E_{\mu_0^k} \big[ \ind_{\big(\log(f_0^k(\tX^k)/f_1^k(\tX^k))\geq \varepsilon\big)} \frac{f_1^k(\tX^k)}{f_0^k(\tX^k) } \big] \nonumber\\
		&=& \E_{\mu_0^k} \Big[\ind_{\big(\log(f_0^k(\tX^k)/f_1^k(\tX^k))\geq \varepsilon\big)} \Big(1-e^\varepsilon \frac{f_1^k(\tX^k)}{f_0^k(\tX^k)} \Big) \Big] \nonumber \\
		&\leq& \E [e^{\lambda Z- \lambda\varepsilon+\lambda\log\lambda-(\lambda+1)\log(\lambda+1)}] \;, \label{eq:deltabound2Gaussian}
\end{eqnarray}
where $\tX^k$ is a random vector distributed  according to $\mu_0^k$,
$Z \equiv \log(f_0^k(\tX^k) / f_1^k(\tX^k))$,
and the last line follows from
$\ind_{(x\geq\varepsilon)}(1-e^{\varepsilon -x}) \leq e^{\lambda(x-\varepsilon)+\lambda\log\lambda-(\lambda+1)\log(\lambda+1)}$ for any $\lambda\geq0$.

Next, we give an upper bound on the moment generating function of $Z$.
\begin{eqnarray*}
	\E_{\mu_0}[e^{\lambda \log(f_0(X)/f_1(X))}] &=& \E[e^{-\lambda\Delta X/\sigma^2}]e^{ \lambda\Delta^2/2\sigma^2} \\
		&\leq& e^{ (\Delta^2/2\sigma^2)\lambda^2 +(\Delta^2/2\sigma^2)\lambda}\;,
\end{eqnarray*}
for any $\lambda\geq0$.
Substituting this into \eqref{eq:deltabound2Gaussian} with a
choice of $\lambda=\frac{\sigma^2}{k\Delta^2}\big(\varepsilon-\frac{k\Delta^2}{2\sigma^2}\big) $, which
is positive for $\varepsilon>k\Delta^2/2\sigma^2$, we get
\begin{eqnarray*}
	\mu_0^k(B)-e^\varepsilon \mu_1^k(B)  &\leq&
	\exp\Big\{ (k\Delta^2/2\sigma^2)\lambda^2 + (k\Delta^2/2\sigma^2) \lambda -\varepsilon\lambda  +\lambda\log\lambda -(\lambda+1)\log(\lambda+1) \Big\} \\
		&\leq& \frac{1}{1+\frac{\sigma^2}{k\Delta^2}\big(\varepsilon-\frac{k\Delta^2}{2\sigma^2}\big)} \exp
			\Big\{ -\frac{\sigma^2}{2k\Delta^2}\Big(\varepsilon-\frac{k\Delta^2}{2\sigma^2}\Big)^2\Big\} \\
		&\leq& \frac{1}{1+\sqrt{\frac{2\sigma^2}{k\Delta^2}\log(e+\frac{1}{\delta}\sqrt{\frac{k\Delta^2}{\sigma^2}})}}
			\frac{1}{e+\frac{1}{\delta}\sqrt{\frac{k\Delta^2}{\sigma^2}}} \\
		&\leq& \frac{1}{\sqrt{\frac{k\Delta^2}{\sigma^2}}+\sqrt{2 \log(e+(1/\delta)\sqrt{k\Delta^2/\sigma^2})}} \frac{\delta}{e\delta\sqrt{\frac{\sigma^2}{k\Delta^2}}+1} \;,
\end{eqnarray*}
for our choice of $\sigma^2$ such that
$\varepsilon \geq k\Delta^2/(2\sigma^2) + \sqrt{(2k\Delta^2/\sigma^2)\log(e+(1/\delta)\sqrt{k\Delta^2/\sigma^2})}$.
The right-hand side is always less than $\delta$.

With $\sigma^2 \geq (4k\Delta^2/\varepsilon^2) \log(e+(\varepsilon/\delta))$
and $\sigma^2 \geq k\Delta^2/(4\varepsilon)$, this ensures that the above condition is satisfied.
This implies that we only need $\sigma^2=O((k\Delta^2/\varepsilon^2) \log(e+(\varepsilon/\delta)))$.
%
%
\section{Analysis of the geometric mechanism in Theorem \ref{thm:geometric}}
\label{sec:geometricproof}

Theorem \ref{thm:geometric} follows directly from the proof of Theorem \ref{thm:composition}, once the appropriate associations are made.
Consider two databases $D_0$ and $D_1$, and a single query $q$ such that
$q(D_1)=q(D_0)+1$.
The geometric mechanism produces two random outputs
$q(D_0)+Z$ and $q(D_1)+Z$ where $Z$ is distributed accruing to the geometric distribution.
Let $P_0$ and $P_1$  denote the distributions of the random output respectively.
For $x\leq q(D_0)$, $P_0(x) = e^\varepsilon P_1(x)$, and
for $x> q(D_0)$, $e^\varepsilon P_0(x) = P_1(x)$.
Then, it is not difficult to see that
the privacy region achieved by the geometric mechanism is equal to the privacy region achieved by
the canonical binary example of $\tX_0$ and $\tX_1$ in \eqref{eq:dist0} and \eqref{eq:dist1} with $\delta=0$.
This follows from the fact there is a stochastic transition from the pair $\tX_0$ and $\tX_1$ to
$q(D_0)+Z$ and $q(D_1)+Z$; further, the converse is also true.
Hence, from the perspective of hypothesis testing, those two (pairs of) outcomes are equivalent.

It now follows from the proof of Theorem~\ref{thm:composition} that the $k$-fold composition privacy region
is exactly the optimal privacy region described in \eqref{eq:composition} with $\delta=0$.
We also know that this is the largest possible privacy region achieved by a class of $(\varepsilon,0)$-differentially private mechanisms.

%
%

\subsection{Cut queries of a graph and variance queries of a matrix}
\label{sec:Application1}

Blocki et.\ al.\ \cite{BBDS12} showed that
classical Johnson-Lindenstrauss transform can be used
to produce a differentially private version of a database.
Further, they show that this achieves the best tradeoff between privacy and utility
for two applications: cut queries of a graph and variance queries of a matrix.
In this section, we show how the best known trade off can be further improved
by applying Theorem \ref{thm:closedform}.


First, Blocki et.\ al.\ provide  a differentially private mechanism for cut queries $q(G,S)$:
the number of edges crossing a ($S,\bS$)-cut in a weighted undirected graph $G$.
This mechanism produces a sanitized graph satisfying $(\varepsilon,\delta)$-differential privacy,
where two graphs are neighbors if they only differ on a single edge.
The {\em utility} of the mechanism is measured
via the additive error $\tau$ incurred by the privatization.
Precisely, a mechanism $M$ is said to give a $(\eta,\tau,\nu)$-approximation for
a {\em single} cut query $q(\cdot,\cdot)$,
if for every graph $G$ and every nonempty $S$ it holds that
\begin{eqnarray}
	\prob\Big(\, (1-\eta)\, q(G,S) - \tau \,\leq\, M(G,S) \,\leq\, (1+\eta)\,q(G,S)+\tau  \,\Big) &\geq& 1-\nu \;.
	\label{eq:JLutil}
\end{eqnarray}

For the proposed Johnson-Lindenstrauss mechanism satisfying $(\varepsilon,\delta$)-differential privacy, it is shown that the additive error $\tau_0$
incurred by querying the database $k$ times is bounded by \cite[Theorem 3.2]{BBDS12}\footnote{The original theorem is stated for a single query with $k=1$. Here we state
it more generally with arbitrary $k$.
This requires scaling $\nu$ by $1/k$ to take into account the
union bound over $k$ query outputs in the {\em utility} guarantee in \eqref{eq:JLutil}.}
\begin{eqnarray}
	\tau_0 &=& O\Big(|S| \frac{ \sqrt{\log(1/\delta)\log(k/\nu)}}{\varepsilon} \, \log \Big(\frac{\log(k/\nu)}{\eta^2 \delta}\Big) \, \Big)\;. \label{eq:JL}
\end{eqnarray}
Compared to
other state-of-the-art privacy mechanisms such as the
Laplace noise adding mechanism \cite{Dwo06},
Exponential mechanism \cite{MT07},
Multiplicative weights \cite{HR10},
and Iterative Database Construction \cite{GRU12},
it is shown in \cite{BBDS12} that
the Johnson-Lindenstrauss mechanism
achieves the best tradeoff between the additive error $\tau_0$ and the
privacy $\varepsilon$.
This tradeoff in \eqref{eq:JL} is proved
using the existing Theorem \ref{thm:boosting}.
We can improve this analysis using the optimal composition theorem of
Theorem \ref{thm:closedform}, which gives
\begin{eqnarray}
	\tau &=& O\Big(|S| \frac{ \sqrt{\log(e+\varepsilon/\delta)\log(k/\nu)}}{\varepsilon} \, \log \Big(\frac{\log(k/\nu)}{\eta^2 \delta}\Big) \, \Big)\;.
\label{eq:JLimprove}
\end{eqnarray}
This is smaller than \eqref{eq:JL} by (a square root of) a logarithmic factor
when $\varepsilon=\Theta(\delta)$.
The proof of the analysis in \eqref{eq:JLimprove} is provided below.

%

A similar technique has been used in \cite{BBDS12} to provide a differentially private
mechanism for variance queries $v(A,x) = x^TA ^TAx$:
the variance of a given matrix in a direction $x$.
The proposed mechanism produces a sanitized covariance matrix that satisfy
$(\varepsilon,\delta)$-differential privacy, where two matrices are neighbors if
they differ only in a single row and the difference is
by Euclidean distance at most one. With the previous composition theorem in Theorem \ref{thm:boosting},
the authors of \cite{BBDS12} get an error bound
$	\tau_1 = O\Big(\, \frac{\log(1/\delta)\log(k/\nu)}{\varepsilon^2\eta}
	\log^2\Big(\frac{\log(k/\nu)}{\eta^2\delta}\Big) \,\Big).$
Using our tight composition theorem, this can be improved as
	$\tau = O\Big(\, \frac{\log(e+\varepsilon/\delta)\log(k/\nu)}{\varepsilon^2\eta}
	\log^2\Big(\frac{\log(k/\nu)}{\eta^2\delta}\Big) \,\Big).$
Again, for $\varepsilon=\Theta(\delta)$, this is an improvement of a logarithmic factor.

For  cut queries, Johnson-Lindenstrauss mechanism proceeds as follows:

\begin{center}
\begin{tabular}{ll}
\hline
\vspace{-.35cm}\\
\multicolumn{2}{l}{ Johnson-Lindenstrauss mechanism for cut queries \cite{BBDS12} }\\
\hline
\vspace{-.35cm}\\
\multicolumn{2}{l}{{\bf Input:} A $n$-node graph $G$, parameters $\varepsilon,\delta,\eta,\nu>0$} \\
\multicolumn{2}{l}{{\bf Output:} An approximate Laplacian of $G$: $\tL$}\\
1:  & Set $r=8\log(2/\nu)/\nu^2$ and $w=\sqrt{32r\log(2/\delta)}\log(4r/\delta)/\varepsilon$\\
2:  & For every pair of nodes $I\neq j$, set new weights $w_{i,j} = w/n + (1-w/n)w_{i,j}$\\
3:  & Randomly draw a matrix $N$ of size $r\times {n \choose 2}$, whose entries are i.i.d. samples of   ${\cal N}(0,1)$\\
4:  & Output $\tL = (1/r) E_G^T N^T N E_G$, \\
& 
where $E_G$ is an ${n \choose 2}\times n$ matrix whose $(i,j)$-th row is $\sqrt{w_{i,j}} (e_i - e_j)$ \\
\hline
\end{tabular}
\end{center}
Here $e_i$ is the standard basis vector with one in the $i$-th entry.
Given this synopsis of the sanitized graph Laplacian, a cut query $q(G,S)$ returns
$1/(1-w/n)(\ones_S^T \tL \ones_S - w |S|(n-|S|)/n)$,
where $\ones_S\in\{0,1\}^n$ is the indicator vector for the set $S$.
If the matrix $N$ is an identity matrix, this returns the correct cut value of $G$.

We have the choice of $w\in\reals$ and $r\in \Z$ to ensure that
the resulting mechanism is $(\varepsilon,\delta)$-differentially private, and
satisfy $(\eta,\tau,\nu)$-approximation guarantees of \eqref{eq:JLutil}.
We utilize the following lemma from \cite{BBDS12}.
\begin{lemma}
	With the choice of
	\begin{eqnarray*}
		w &=& \frac{4}{\varepsilon_0} \log(2/\delta_0) \;\;\;\text{ and }\;\;\; r \;=\; \frac{8\log(2/\nu)}{\eta^2}\;,
	\end{eqnarray*}
	each row of $N E_G$ satisfy $(\varepsilon_0,\delta_0)$-differential privacy, and
	the resulting Johnson-Lindenstrauss mechanism satisfy $(\eta,\tau,\nu)$-approximation
	guarantee with
	\begin{eqnarray*}
		\tau &=& 2|S|\,\eta\, w\;,
	\end{eqnarray*}
	where $|S|$ is the size of the smaller partition $S$ of the cut $(S,\bS)$.
\end{lemma}
The error bound in \eqref{eq:JL} follows from choosing
\begin{eqnarray*}
	\varepsilon_0 &=& \frac{\varepsilon}{\sqrt{4r \log(2/\delta)}} \;\;\;\text{ and } \;\;\; \delta_0 \;=\; \frac{\delta}{2 r}\;,
\end{eqnarray*}
and applying Theorem \ref{thm:boosting} to
 ensure that the resulting mechanism with $r$-composition of the $r$ rows of
$M E_G$ is $(\varepsilon,\delta)$-differentially private.
Here it is assumed that $\varepsilon<1$.

Now, with Theorem \ref{thm:composition},
we do not require $\varepsilon_0$ to be as small,
which in turn allows us to add smaller noise $w$,
giving us an improved error bound on $\tau$.
Precisely, using Theorem \ref{thm:closedform} it follows that a choice of
\begin{eqnarray*}
	\varepsilon_0 &=& \frac{\varepsilon}{\sqrt{4r \log(e+2\varepsilon/\delta)}}  \;\;\;\text{ and } \;\;\; \delta_0 \;=\; \frac{\delta}{2 r}\;,
\end{eqnarray*}
suffices to ensure that after $r$-composition we get
$(\varepsilon,\delta)$-differential privacy.
Resulting noise is bounded by
$w \leq 4\sqrt{4r\log(e+2\varepsilon/\delta)}\log(4r/\delta)/\varepsilon$,
which gives the error bound in \eqref{eq:JLimprove}.
The proof follows analogously for the matrix variance queries.

\section{Proof of Theorem \ref{thm:average}}
\label{proof_multi_party}
To prove Theorem \ref{thm:average},
it is sufficient to prove Theorem \ref{thm:proofaverage} stating that
that any other protocol can be simulated from the randomized response outputs.
Let $\{x_i\}_{i\in[k]}$ denote the binary data distributed among $k$ parties.
Let $\{\tx_i\}_{i\in[k]}$ denote the outputs of the randomized response as per Equation \eqref{eq:rr}.
We will prove that any protocol that obeys differential privacy conditions  can be simulated from $\tx_i$'s.
This proves the desired theorem,
since the optimal protocol and the optimal decision rules can be simulated by
each node (and the central observer) upon receiving the randomized responses.
Hence, proving that randomized response is sufficient to achieve optimal performance (on any metric).

\begin{thm}
	\label{thm:proofaverage}
	For any protocol that generates a random transcript $\tau$,
	there exists a stochastic transformation $T$ such that the joint distribution of the bits and the transcript can be
	simulated from the randomized outputs:
	\begin{eqnarray}
		(x_1,\ldots,x_k,\tau) &\stackrel{D}{=}& (x_1,\ldots,x_k,T(\tx_1,\ldots,\tx_k))\;, \label{eq:proofaverage}
	\end{eqnarray}
	where $\stackrel{D}{=}$ denotes equality in distribution, and $\tx_i$ is a randomized response of $x_i$.
\end{thm}

To prove the above theorem, our strategy is to apply induction argument over a class of stochastic transformations
$T_\ell$ taking randomized responses $\tx_1^\ell=(\tx_1,\ldots,\tx_\ell)$ as inputs together with the original bits
$x_{\ell+1}^k=(x_{\ell+1},\ldots,x_k)$. We will prove the following series of equations:
\begin{eqnarray}
	(x_1,\ldots,x_k,\tau) &\stackrel{D}{=}& (x_1,\ldots,x_k,T_1(\tx_1, x_2^k )) \label{eq:proofave1}\\
		 &\stackrel{D}{=}& (x_1,\ldots,x_k,T_2(\tx_1^2,x_3^k)) \label{eq:proofave2}\\
		 &\vdots&  \nonumber\\
		 &\stackrel{D}{=}& (x_1,\ldots,x_k,T_k(\tx_1^k))  \;, \label{eq:proofave3}
\end{eqnarray}

We first prove Equation \eqref{eq:proofave1}.
We show an equivalent version of this equation, which is
$ (x_1,\tau) \stackrel{D}{=} (x_1,T(\tx_1,x_2^k))$ for all fixed values of $x_2^k$.
Equation \eqref{eq:proofave1} follows by applying Bayes rule to this equation.
First, note that for all fixed $x_2^k$,
\begin{eqnarray}
	\cR\big({\tau },x_1=0,x_1=1 \big) &\subseteq& \cR(\varepsilon_1,\delta_1)\;,\label{eq:regionbound1}
\end{eqnarray}
by the fact that $\tau$ is $(\varepsilon_1,\delta_1)$-differentially private and Corollary \ref{coro:hypo}.
Next, notice that by construction, the randomized response achieves this outer bound, i.e.
\begin{eqnarray}
	\cR\big(\tx_1,x_1=0,x_1=1 \big) &=& \cR(\varepsilon_1,\delta_1)\;, \label{eq:regionbound2}
\end{eqnarray}
for all values of $x_2^k$ which holds only under the current assumption that $x_1^k$ are independent.
Hence from the reverse data processing inequality in Theorem \ref{thm:converse},
it follows that for each instance of $x_2^k$, there exists a stochastic transformation
such that $\tau$ is simulated from $\tx_1$, i.e. $ (x_1,\tau) \stackrel{D}{=} (x_1,T(\tx_1,x_2^k))$.
This proves the desired Equation \eqref{eq:proofave1}.

Now, we prove the induction step that
starting from Equation \eqref{eq:proofave1} allows us to show recursively
Equations \eqref{eq:proofave2} and \eqref{eq:proofave3}.
We want to prove that there always exists a stochastic transformation $T_{\ell+1}$ such that
\begin{eqnarray}
	(x_1^k,T_\ell(\tx_1^\ell, x_{\ell+1}^k )) & \stackrel{D}{=} & (x_1^k,T_{\ell+1}(\tx_1^{\ell+1}, x_{\ell+2}^k )) \;,
	\label{eq:proofave5}
\end{eqnarray}
for any stochastic transformation $T_\ell$ satisfying $(\varepsilon_{\ell+1},\delta_{\ell+1})$-differential privacy.
Again, we   prove that
$(x_{\ell+1},T_\ell(\tx_1^\ell, x_{\ell+1}^k )) \stackrel{D}{=} (x_{\ell+1} ,T_{\ell+1}(\tx_1^{\ell+1}, x_{\ell+2}^k ))$
for all values of $(x_1^\ell,\tx_1^\ell,x_{\ell+1}^k)$. Then, Equation \eqref{eq:proofave5} follows from Bayes rule.
First note that from the assumption that $T_\ell(\tx_1^\ell,x_{\ell+1}^k)$ is $(\varepsilon_{\ell+1},\delta_{\ell+1})$-differentially private
with respect to $x_{\ell+1}$, we know that
for any fixed values of $(x_1^\ell,\tx_1^\ell,x_{\ell+2}^k)$,
binary hypothesis testing on $x_{\ell+1}$ based on the observation $T_\ell(\tx_1^\ell,x_{\ell+1}^k)$ must obey the differential privacy constraint:
\begin{eqnarray}
	\prob( T_\ell(\tx_1^\ell,x_{\ell+1}^k)\in S | x_{\ell+1}, x_1^\ell, \tx_1^\ell,x_{\ell+2}^k) &\leq& e^{\varepsilon_{\ell+1}} \prob(T_\ell(\tx_1^\ell,x_{\ell+1}^k)\in S | \overline{x_{\ell+1}}, x_1^\ell,\tx_1^\ell, x_{\ell+2}^k)+ \delta_{\ell+1}\;,
\end{eqnarray}
and since $T_\ell(\tx_1^\ell,x_{\ell+1}^k)$ is conditionally independent of
$x_1^\ell$ given $\tx_1^\ell$, we get
\begin{eqnarray}
	\prob( T_\ell(\tx_1^\ell,x_{\ell+1}^k)\in S | x_{\ell+1}, \tx_1^\ell,x_{\ell+2}^k) &\leq& e^{\varepsilon_{\ell+1}} \prob(T_\ell(\tx_1^\ell,x_{\ell+1}^k)\in S | \overline{x_{\ell+1}}, \tx_1^\ell, x_{\ell+2}^k)+ \delta_{\ell+1}\;.
\end{eqnarray}
This implies that for each value of $(\tx_1^\ell,x_{\ell+2}^k)$,
\begin{eqnarray}
	\cR\big(T_\ell(\tx_1^\ell,x_{\ell+1}^k),x_{\ell+1}=0,x_{\ell+1}=1 \big) &\subseteq& \cR(\varepsilon_{\ell+1},\delta_{\ell+1})\;. \label{eq:regionbound5}
\end{eqnarray}
 Next, notice that by construction, the randomized response achieves this outer bound, i.e.
\begin{eqnarray}
	\cR\big(\tx_{\ell+1},x_1=0,x_1=1 \big) &=& \cR(\varepsilon_{\ell+1},\delta_{\ell+1})\;, \label{eq:regionbound6}
\end{eqnarray}
for all values of $(\tx_1^\ell,x_{\ell+2}^k)$ which holds only under the current assumption that $x_1^k$ are independent.
Hence from the reverse data processing inequality in Theorem \ref{thm:converse},
it follows that for each instance of $(\tx_1^\ell,x_{\ell+2}^k)$, there exists a stochastic transformation
such that $T_\ell$ is simulated from $\tx_{\ell+1}$, i.e. $ (x_{\ell+1},T_\ell(\tx_1^\ell,x_{\ell+1}^k)) \stackrel{D}{=} (x_{\ell+1}, T_{\ell+1}(\tx_{\ell+1},\tx_1^\ell,x_{\ell+2}^k))$.
This proves the desired induction step in Equation \eqref{eq:proofave5}.
Consequently, by induction Equation \eqref{eq:proofave3} holds, and this proves desired Theorem \ref{thm:proofaverage}.

\bibliographystyle{amsalpha}

\bibliography{privacy,privacy2,references}

\newcommand{\etalchar}[1]{$^{#1}$}
\providecommand{\bysame}{\leavevmode\hbox to3em{\hrulefill}\thinspace}
\providecommand{\MR}{\relax\ifhmode\unskip\space\fi MR }
\providecommand{\MRhref}[2]{%
  \href{http://www.ams.org/mathscinet-getitem?mr=#1}{#2}
}
\providecommand{\href}[2]{#2}
\begin{thebibliography}{DKM{\etalchar{+}}06b}

\bibitem[AS04]{AS04}
Noga Alon and Joel~H Spencer, \emph{The probabilistic method}, Wiley. com,
  2004.

\bibitem[BBDS12]{BBDS12}
Jeremiah Blocki, Avrim Blum, Anupam Datta, and Or~Sheffet, \emph{The
  johnson-lindenstrauss transform itself preserves differential privacy},
  Foundations of Computer Science (FOCS), 2012 IEEE 53rd Annual Symposium on,
  IEEE, 2012, pp.~410--419.

\bibitem[BDMN05]{BDMN05}
Avrim Blum, Cynthia Dwork, Frank McSherry, and Kobbi Nissim, \emph{Practical
  privacy: the {SuLQ} framework}, Proceedings of the twenty-fourth ACM
  SIGMOD-SIGACT-SIGART symposium on Principles of database systems, ACM, 2005,
  pp.~128--138.

\bibitem[Bla53]{Bla53}
David Blackwell, \emph{Equivalent comparisons of experiments}, The Annals of
  Mathematical Statistics \textbf{24} (1953), no.~2, 265--272.

\bibitem[Bla65]{Bla65}
N~Blachman, \emph{The convolution inequality for entropy powers}, Information
  Theory, IEEE Transactions on \textbf{11} (1965), no.~2, 267--271.

\bibitem[BLR13]{BLR13}
Avrim Blum, Katrina Ligett, and Aaron Roth, \emph{A learning theory approach to
  noninteractive database privacy}, Journal of the ACM (JACM) \textbf{60}
  (2013), no.~2, 12.

\bibitem[BNO08]{BNO08}
Amos Beimel, Kobbi Nissim, and Eran Omri, \emph{Distributed private data
  analysis: Simultaneously solving how and what}, Advances in
  Cryptology--CRYPTO 2008, Springer, 2008, pp.~451--468.

\bibitem[CT88]{CT92}
Thomas~M Cover and A~Thomas, \emph{Determinant inequalities via information
  theory}, SIAM journal on Matrix Analysis and Applications \textbf{9} (1988),
  no.~3, 384--392.

\bibitem[CT12]{CT06}
Thomas~M Cover and Joy~A Thomas, \emph{Elements of information theory}, John
  Wiley \& Sons, 2012.

\bibitem[DCT91]{DCT91}
Amir Dembo, Thomas~M Cover, and Joy~A Thomas, \emph{Information theoretic
  inequalities}, Information Theory, IEEE Transactions on \textbf{37} (1991),
  no.~6, 1501--1518.

\bibitem[DJW13]{DJW13}
J.~C. Duchi, M.~I. Jordan, and M.~J. Wainwright, \emph{Local privacy and
  statistical minimax rates}, Foundations of Computer Science (FOCS), 2013 IEEE
  54th Annual Symposium on, IEEE, 2013, pp.~429--438.

\bibitem[DKM{\etalchar{+}}06a]{DKM06}
Cynthia Dwork, Krishnaram Kenthapadi, Frank McSherry, Ilya Mironov, and Moni
  Naor, \emph{Our data, ourselves: Privacy via distributed noise generation},
  Advances in Cryptology-EUROCRYPT 2006, Springer, 2006, pp.~486--503.

\bibitem[DKM{\etalchar{+}}06b]{DKMMN}
\bysame, \emph{Our data, ourselves: Privacy via distributed noise generation},
  Advances in Cryptology-EUROCRYPT 2006, Springer, 2006, pp.~486--503.

\bibitem[DL09]{DL09}
C.~Dwork and J.~Lei, \emph{Differential privacy and robust statistics},
  Proceedings of the 41st annual ACM symposium on Theory of computing, ACM,
  2009, pp.~371--380.

\bibitem[DMNS06]{DMNS06}
Cynthia Dwork, Frank McSherry, Kobbi Nissim, and Adam Smith, \emph{Calibrating
  noise to sensitivity in private data analysis}, Theory of Cryptography,
  Springer, 2006, pp.~265--284.

\bibitem[DN03]{DN03}
Irit Dinur and Kobbi Nissim, \emph{Revealing information while preserving
  privacy}, Proceedings of the twenty-second ACM SIGMOD-SIGACT-SIGART symposium
  on Principles of database systems, ACM, 2003, pp.~202--210.

\bibitem[DN04]{DN04}
Cynthia Dwork and Kobbi Nissim, \emph{Privacy-preserving datamining on
  vertically partitioned databases}, Advances in Cryptology--CRYPTO 2004,
  Springer, 2004, pp.~528--544.

\bibitem[DRV10]{DRV10}
Cynthia Dwork, Guy~N Rothblum, and Salil Vadhan, \emph{Boosting and
  differential privacy}, Foundations of Computer Science (FOCS), 2010 51st
  Annual IEEE Symposium on, IEEE, 2010, pp.~51--60.

\bibitem[Dwo06]{Dwo06}
Cynthia Dwork, \emph{Differential privacy}, Automata, languages and
  programming, Springer, 2006, pp.~1--12.

\bibitem[GMPS13]{GMPS}
Vipul Goyal, Ilya Mironov, Omkant Pandey, and Amit Sahai,
  \emph{Accuracy-privacy tradeoffs for two-party differentially private
  protocols}, Advances in Cryptology--CRYPTO 2013, Springer, 2013,
  pp.~298--315.

\bibitem[GRS12]{GRS12}
Arpita Ghosh, Tim Roughgarden, and Mukund Sundararajan, \emph{Universally
  utility-maximizing privacy mechanisms}, SIAM Journal on Computing \textbf{41}
  (2012), no.~6, 1673--1693.

\bibitem[GRU12]{GRU12}
Anupam Gupta, Aaron Roth, and Jonathan Ullman, \emph{Iterative constructions
  and private data release}, Theory of Cryptography, Springer, 2012,
  pp.~339--356.

\bibitem[GV12]{GV12}
Quan Geng and Pramod Viswanath, \emph{Optimal noise-adding mechanism in
  differential privacy}, arXiv preprint arXiv:1212.1186 (2012).

\bibitem[GV13]{GV13}
\bysame, \emph{The optimal mechanism in $(\epsilon,\delta)$-differential
  privacy}, arXiv preprint arXiv:1305.1330 (2013).

\bibitem[HLM10]{HLM10}
Moritz Hardt, Katrina Ligett, and Frank McSherry, \emph{A simple and practical
  algorithm for differentially private data release}, arXiv preprint
  arXiv:1012.4763 (2010).

\bibitem[HR10]{HR10}
Moritz Hardt and Guy~N Rothblum, \emph{A multiplicative weights mechanism for
  privacy-preserving data analysis}, Foundations of Computer Science (FOCS),
  2010 51st Annual IEEE Symposium on, IEEE, 2010, pp.~61--70.

\bibitem[HR13]{HR13}
Moritz Hardt and Aaron Roth, \emph{Beyond worst-case analysis in private
  singular vector computation}, Proceedings of the 45th annual ACM symposium on
  Symposium on theory of computing, ACM, 2013, pp.~331--340.

\bibitem[HT10]{HT10}
Moritz Hardt and Kunal Talwar, \emph{On the geometry of differential privacy},
  Proceedings of the 42nd ACM symposium on Theory of computing, ACM, 2010,
  pp.~705--714.

\bibitem[KOV14a]{kov14-0}
P.~Kairouz, S.~Oh, and P.~Viswanath, \emph{The composition theorem for
  differential privacy}, International Conference on Machine Learning, 2014.

\bibitem[KOV14b]{KOV14-1}
\bysame, \emph{Extremal mechanisms for local differential privacy}, Advances in
  Neural Information Processing Systems 27, 2014, pp.~2879--2887.

\bibitem[KOV15]{KOV15}
\bysame, \emph{Secure multi-party differential privacy}, Advances in Neural
  Information Processing Systems, 2015.

\bibitem[Lau96]{Lau96}
S.~L. Lauritzen, \emph{{Graphical Models}}, Oxford University Press, 1996.

\bibitem[LV07]{LV07}
Tie Liu and Pramod Viswanath, \emph{An extremal inequality motivated by
  multiterminal information-theoretic problems}, Information Theory, IEEE
  Transactions on \textbf{53} (2007), no.~5, 1839--1851.

\bibitem[MMP{\etalchar{+}}10]{MMPT}
Andrew McGregor, Ilya Mironov, Toniann Pitassi, Omer Reingold, Kunal Talwar,
  and Salil Vadhan, \emph{The limits of two-party differential privacy},
  Foundations of Computer Science (FOCS), 2010 51st Annual IEEE Symposium on,
  IEEE, 2010, pp.~81--90.

\bibitem[MN12]{MN12}
S~Muthukrishnan and Aleksandar Nikolov, \emph{Optimal private halfspace
  counting via discrepancy}, Proceedings of the 44th symposium on Theory of
  Computing, ACM, 2012, pp.~1285--1292.

\bibitem[MT07]{MT07}
Frank McSherry and Kunal Talwar, \emph{Mechanism design via differential
  privacy}, Foundations of Computer Science, 2007. FOCS'07. 48th Annual IEEE
  Symposium on, IEEE, 2007, pp.~94--103.

\bibitem[Sta59]{Sta59}
AJ~Stam, \emph{Some inequalities satisfied by the quantities of information of
  fisher and shannon}, Information and Control \textbf{2} (1959), no.~2,
  101--112.

\bibitem[VG06]{VG06}
Sergio Verd{\'u} and Dongning Guo, \emph{A simple proof of the entropy-power
  inequality}, IEEE Transactions on Information Theory \textbf{52} (2006),
  no.~5, 2165--2166.

\bibitem[War65]{War65}
S.~L. Warner, \emph{Randomized response: A survey technique for eliminating
  evasive answer bias}, Journal of the American Statistical Association
  \textbf{60} (1965), no.~309, 63--69.

\bibitem[WZ10]{WZ10}
Larry Wasserman and Shuheng Zhou, \emph{A statistical framework for
  differential privacy}, Journal of the American Statistical Association
  \textbf{105} (2010), no.~489, 375--389.

\bibitem[Zam98]{Z98}
Ram Zamir, \emph{A proof of the fisher information inequality via a data
  processing argument}, Information Theory, IEEE Transactions on \textbf{44}
  (1998), no.~3, 1246--1250.

\end{thebibliography}

\end{document}